\documentclass[onecolumn,epjc3]{svjour3}  

\smartqed  % flush right qed marks, e.g. at end of proof
\RequirePackage{graphicx}
\RequirePackage{color}
 \RequirePackage{mathptmx}      % use Times fonts if available on your TeX system
\RequirePackage{latexsym}
\RequirePackage[numbers,sort&compress]{natbib}
\RequirePackage[colorlinks,citecolor=blue,urlcolor=blue,linkcolor=blue]{hyperref}
\RequirePackage{hyperref}
\RequirePackage{dsserif}
\RequirePackage[normalem]{ulem}

\journalname{Eur. Phys. J. Plus }

\begin{document}
\title{Quasi-one- and quasi-two-dimensional Bose-Fermi mixtures from weak coupling to unitarity}

\titlerunning{Bose-Fermi mixture from weak coupling to unitarity}

\author{Pardeep Kaur\footnote{2018phz0004@iitrpr.ac.in}
        \and
        Sandeep Gautam\footnote{sandeep@iitrpr.ac.in}
        \and\\
        S. K. Adhikari\footnote{sk.adhikari@unesp.br}
         %etc.
}

%\footnote{e1}{e-mail:  2018phz0004@iitrpr.ac.in}
%\thankstext{e2}{e-mail:  sandeep@iitrpr.ac.in}
%\thankstext{e3}{e-mail: sk.adhikari@unesp.br}
%\thankstext{e1}{e-mail: fauthor@example.com}


\institute{$^*$Department of Physics, Indian Institute of Technology Ropar, Rupnagar, Punjab 140001, India,
 \label{addr1}
           \and
           $^{**}$Department of Physics, Indian Institute of Technology Ropar, Rupnagar, Punjab 140001, India,
 \label{addr2}
           $^{***}$Instituto de F\'{\i}sica Te\'orica, UNESP - Universidade Estadual Paulista, 01.140-070 S\~ao Paulo, S\~ao Paulo, Brazil. 
           \label{addr3}
 %          \and
  %         \emph{Present Address:} if needed\label{addr3}
}

\date{Received: date / Accepted: date}

\maketitle

\begin{abstract}

We study ultracold superfluid Bose-Fermi mixtures in three dimensions, with stronger
confinement along one or two directions, using a non-perturbative beyond-mean-field 
model for bulk chemical potential valid along the weak-coupling to unitarity crossover.
Although bosons are considered to be in a superfluid state, we consider two 
possibilities for the fermions $-$ spin-polarized degenerate state and superfluid state.
Simplified reduced analytic lower-dimensional models are derived along the weak-coupling
to unitarity  \, crossover in quasi-one-dimensional (quasi-1D) and quasi-two-dimensional
(quasi-2D) settings. The only parameters in these models are the constants of the 
beyond-mean-field Bose-Bose and Fermi-Fermi Lee-Huang-Yang interactions and the
respective universal Bertsch parameter at unitarity. In addition to the numerical
results for a fully-trapped system, we also present results for quasi-2D Bose-Fermi
mixtures where one of the components is untrapped but localized due to the interaction
mediated by the other component. We demonstrate the validity of the reduced quasi-1D
and quasi-2D models via a comparison of the numerical solutions for the ground states 
obtained from the reduced models and the full three-dimensional (3D) model.
  
\end{abstract}

%\pacs{03.75.Lm,03.75.Kk, 05.45.Yv}

\section {Introduction} 
\label{I}
For the last two decades or so, both experimental \cite{Binary-expt1,Binary-expt2,
Binary-expt3,Binary-expt4} and theoretical \cite{Binary-theo1,Binary-theo2,Binary-theo3}
investigations have been undertaken on the two-component mixtures of quantum degenerate 
atomic gases. While the initial studies mainly focused on the degenerate Bose-Bose mixtures,
later with the experimental realization of stable quantum degenerate Bose-Fermi mixtures {consisting
of a superfluid Bose gas and a spin-polarized degenerate Fermi gas}
\cite{Bosonfermi-expt1, Bosonfermi-expt2, Bosonfermi-expt5, Bosonfermi-expt3, Chin_expt_BFsoliton, 
Bosonfermi-expt4, bfex1,bfex2,bfex3,bfex4, Bosonfermi-expt6, Bosonfermi-expt7}, these systems
attracted particular interest. The quantum degenerate Bose-Fermi mixtures of $^{87}$Rb-$^{40}$K 
\cite{Bosonfermi-expt1,Bosonfermi-expt2,bfex1,bfex2,bfex3}, $^{23}$Na-$^{6}$Li \cite{bfex4},
$^{7}$Li-$^{6}$Li \cite{Bosonfermi-expt7}, $^{87}$Rb-$^{171}$Yb \cite{Bosonfermi-expt4}, 
$^{41}$K-$^{6}$Li \cite{Bosonfermi-expt5}, $^{174}$Yb-$^{6}$Li \cite{Bosonfermi-expt6}, $^{133}$Cs-$^{6}$Li 
\cite{Bosonfermi-expt3,Chin_expt_BFsoliton}, etc. are among the experimental realizations of such systems.
These Bose-Fermi quantum degenerate mixtures have mass imbalance ranging from a small to a large value.
With a lighter fermionic component ($^6${Li}) and a heavier bosonic component ($^{133}${Cs}), 
$^{133}$Cs-$^{6}$Li quantum degenerate mixture has a large mass imbalance that offers rich interaction
properties that are well characterized \cite{pro-cs-li,pro-cs-li-2,pro-cs-li-3,pro-cs-li-4,pro-cs-li-5}.
In a recent experiment Desalvo {\em{et al.}} \cite{Chin_expt_BFsoliton} showed that in a Bose-Einstein
condensate (BEC) of caesium atoms inside a {spin-polarized} quantum degenerate Fermi gas of lithium atoms, interspecies
interactions can lead to an effective trapping potential via a fermion-mediated attractive boson-boson
interactions. Earlier in another experiment, Desalvo {\em{et al.}} \cite{Bosonfermi-expt3} had showed
that a {spin-polarized} degenerate Fermi gas of $^{6}$Li, even without an external trapping, can be fully confined in
a harmonically trapped $^{133}$Cs BEC. These degenerate Bose-Fermi mixtures also paved the way to
realize self-trapped entities termed as soliton trains \cite{Gajda_PRL_BFsoliton,Chin_expt_BFsoliton,adhisol}.

To study a superfluid Bose-Fermi mixture, the mean-field treatment is commonly used, which accounts for the weakly-interacting 
limit of interactions in the mixture, encapsulated by the Gross-Pitaevskii (GP) equation for bosons \cite{gross,gross1,lagboson}
and  Bardeen-Cooper-Schrieffer (BCS) mean-field equations for fermions \cite{lagfermion}. In the case of fermions, 
simpler superfluid hydrodynamic equations are commonly used to study the physics characterized by wavelengths longer
than the healing length \cite{HYD-eqs,HYD-eqs1}. These superfluid hydrodynamic equations are equivalent to a time-dependent
non-linear \, \, Schr\"odinger equation which can be used to study the macroscopic properties of these weakly-interacting 
superfluids including collective excitations \cite{sch-like-eqn,sch-like-eqn1,sch-like-eqn2,sle0,sch-like-eqn3}. To account for strong interactions, crossover models
have also been proposed in the literature that connect weak-coupling and unitarity limits for bosons 
\cite{boson,BoseFermi_analytical} and fermions \cite{Fermi_sadhan,BoseFermi_analytical,gautam_adhikari}.
The crossover model for fermions enjoyed widespread usage \cite{wenwen,wenwen1,wenwen2,wenwen3,wenwen4}.
Nonetheless, interspecies Bose-Fermi interaction in these studies still corresponds to weakly-interacting domain. 
Lee, Huang, and Yang \cite{LHYbf} (LHY) provided the lowest-order beyond-mean-field correction to energy for the Bose
\cite{LHYb} and  Fermi \cite{LHYf} systems, which  played a vital role in the development of these crossover models. 
Usually, these crossover models \cite{boson,BoseFermi_analytical,Fermi_sadhan} for Bose and Fermi systems produce the
correct weak-coupling LHY as well as the strong-coupling unitarity limits. However, these crossover
models for Bose and Fermi interactions, for describing the superfluid Bose-Fermi mixtures between weak- and  
strong-coupling regimes, usually have fitting parameters. In this study, we propose minimal crossover models
without adjustable fitting parameters to describe the Bose and Fermi interactions from the weak-coupling to unitarity
limits for quasi-one-dimensional (quasi-1D) and quasi-two-dimensional (quasi-2D) configurations.
These are derived from the corresponding three-dimensional (3D) model, by integrating out the dependence
on the transverse coordinates by the usual procedure \cite{sala-gll}. 
The only parameters in these reduced quasi-1D and quasi-2D models are the constants
of the  Bose-Bose and Fermi-Fermi LHY interactions and the universal Bertsch parameter \cite{bertsch} at unitarity.
We also present the relevant equations for a degenerate Bose-Fermi mixture in 3D, quasi-1D and quasi-2D configurations, 
as many of the experiments were performed under these conditions.  In this case the energy of the bosons is described 
by the crossover model varying from weak-coupling to unitarity limit, whereas that of the noninteracting degenerate fermions 
{has the contribution from the Fermi pressure term only}.

Using the analytic weak-coupling to unitary crossover functions for bulk chemical potentials
and energy densities of Bose and Fermi components, we write the 3D beyond mean-field equations
for both a superfluid and a degenerate Bose-Fermi mixture valid along the crossover from the weak-coupling to the 
unitarity limits, from which the reduced quasi-1D and quasi-2D model equations are derived. 
The quasi-1D and quasi-2D traps are of interest as many experiments are performed in these settings.
The resulting coupled non-linear Schr\"odinger equation has bulk chemical potentials in the form of 
analytic Pad\'e approximants  connecting the weak and strong coupling limits. We use these crossover
models to study  Bose-Fermi mixtures confined by quasi-1D and quasi-2D trapping potentials, and compare  
the results of atom densities with those obtained from the 3D crossover model. The Bose-Fermi mixture 
is miscible (immiscible) for weak (strong) Bose-Fermi repulsion \cite{mis-im,mis-im1,mis-im2,mis-im3,mis-im4,mis-im5}. Here we consider miscible 
as well as phase-separated Bose-Fermi mixtures for repulsive Bose-Bose and Bose-Fermi interactions.
In addition to the study of fully-trapped systems with repulsive Bose-Bose and Bose-Fermi interactions,
we also consider the problem of self-trapping of one of the components as a result of the interactions 
mediated by the other component in quasi-2D Bose-Fermi superfluid and degenerate mixtures for attractive Bose-Fermi and 
repulsive Bose-Bose interactions, as explored in recent experiments \cite{Bosonfermi-expt3,Chin_expt_BFsoliton} 
for a degenerate Bose-Fermi mixture.

The following depicts how the paper is arranged. 
The analytical equations for the energy density and bulk chemical potential of 
homogeneous Bose and Fermi superfluids {along with those of spin-polarized degenerate Fermi gas} 
in three dimensions along the weak coupling to unitarity crossover are presented in Sec.~\ref{II}. In Sec. \ref{III}, we derive the
analytic expressions for energy density and bulk chemical potential for bosons in quasi-1D 
and quasi-2D configurations along the weak coupling to unitarity crossover and derive the respective
nonlinear Schr\"odinger equations. In Sec. \ref{IV}, we derive the same for fermions in quasi-1D and
quasi-2D configurations. In Sec. \ref{V}, we derive the crossover model for {superfluid and degenerate} 
Bose-Fermi mixtures in quasi-1D, quasi-2D and 3D configurations in dimensionless units valid along the 
weak-coupling to unitarity limits for intraspecies interactions. The interspecies interaction is
taken to be in the weakly interacting regime. In Sec. \ref{VI}, we first present numerical results 
for phase-separated and miscible quasi-1D and quasi-2D Bose-Fermi superfluid mixtures and
later numerical results for Bose (Fermi) mediated trapping of Fermi (Bose) superfluid component in 
quasi-2D mixtures using the full 3D and reduced models. We also present results for Bose (Fermi) mediated trapping of 
Fermi (Bose) component in quasi-2D degenerate Bose-Fermi mixtures using the full 3D and reduced models. 
In Sec. \ref{VII} we present a brief summary of this study.

\section{Super-fluid hydrodynamics and nonlinear equations}
\label{II}
In the unitarity limit, $|a_i|\to \infty$ resulting in a Bose/Fermi superfluid system of $N_i$ atoms
with a universal behavior \cite{univ,univ1,univ2,univ3,gior,gior1} governed by particle density $n_i$ alone, 
where $a_i$ is the intraspecies scattering length. Here $i = B$ 
for bosons and $i = F$ for paired fermions. 
The bulk chemical potential of a homogeneous Bose or Fermi gas at unitarity 
is 
\cite{BoseFermi_analytical,lagfermion}
\begin{equation}\label{unit}
\lim_{|a_i|\to \infty}\mu_i(n_i,a_i)= \frac{\hbar^2}{m_i}
\eta_i n_i^{2/3},
\end{equation}
where $\eta_i$ is the universal Bertsch  parameter \cite{bertsch}, and $m_i$ is the mass of an atom. 
By dimensional argument Eq. (\ref{unit})  is the simplest expression  with the dimension of energy using particle density $n_i$.

The bulk chemical potential $\mu_B$ of a homogeneous Bose gas in the weak-coupling limit is 
given by \cite{LHYb}
\begin{eqnarray}\label{LHYB}
\mu_B(n_B,a_B) &=&\frac{\hbar^2}{m_B} (4 \pi n_B a_B+2\pi \alpha n_B^{3/2}  
a_B^{5/2}+...),  
\end{eqnarray}
where $\alpha = 64/(3\sqrt \pi$). The first term in this expression for $\mu_B$ is the mean-field GP chemical potential
per particle \cite{gross,gross1} and the second term is the
perturbative LHY correction \cite{LHYb} to it; the latter is meaningful only for small to   moderate 
values of the scattering length $a_B$ $(>0)$. 
The {\it minimal} or the simplest analytic form of the non-perturbative bulk chemical potential along the crossover 
between weak coupling to unitarity is  \cite{gautam_adhikari}
\begin{eqnarray}\label{muB}
\mu_B(n_B,a_B)&\equiv&\frac{\hbar^2}{m_B}  n_B^{2/3}f(\tau),\quad  \tau=a_B n_B   ^ {1/3}, \\
f(\tau) &=&4\pi  \frac{\tau+\alpha \tau^{5/2}}{1+\frac{\alpha}{2}\tau^{3/2}
+\frac{4\pi \alpha}{\eta_B}\tau^{5/2}}, \label{coB}
\end{eqnarray}  
which has the correct unitarity and weak-coupling limits, viz. Eqs. (\ref{unit}) and (\ref{LHYB}), respectively.
Expression (\ref{coB}) is written in a numerator-over-denominator form like a Pad\'e-type approximant 
and  is valid for all values of the scattering length $a_B$ joining the weak-coupling
 and unitarity limits.
The bulk chemical potential $\mu_i$ and energy density ${\cal E}_i$  of a homogeneous Bose 
or Fermi system are related by  
 \begin{eqnarray}\label{mu}
\mu_i(n_i,a_i)= \frac{\partial [n_i{{\cal E}_i(n_i,a_i)}]}{\partial n_i},
\end{eqnarray}
or, equivalently, by
 \begin{eqnarray}\label{EN}
{\cal E}_i(n_i,a_i)= \frac{ 1}{ n_i} \int_0^{n_i} \mu_i(n,a_i)dn  \, .
\end{eqnarray}
Using Eq. (\ref{EN}), the 
bosonic energy density consistent with Eq. (\ref{LHYB}), valid in the weak-coupling limit,  is 
\begin{eqnarray}\label{LHYBE}
{\cal E}_B(n_B,a_B) &    \equiv & \frac{1}{n_B}\int_0^{n_B}  \mu _B(n,a_B) d n\nonumber \\
&=&
\frac{\hbar^2}{m_B} 
\left(2 \pi n_B a_B+\frac{4}{5}\pi \alpha n_B^{3/2}  a_B^{5/2}+...\right),  
\end{eqnarray}
and the same at unitarity, in accordance with Eq. (\ref{unit}), is
\begin{eqnarray}\label{LHYBE2}
{\cal E}_B(n_B,a_B)= \frac{\hbar^2}{m_B} \frac{3}{5} \eta_B n_B^{2/3}.  
\end{eqnarray}
The following Pad\'e-type approximant for the energy density valid
 along the weak-coupling to unitarity crossover \cite{gautam_adhikari} 
\begin{eqnarray} \label{enB}
{\cal E}_B(n_B,a_B)&\equiv&\frac{\hbar^2}{m_B} \frac{2\pi n_B^{2/3}(\tau+\frac{4\alpha}{5} \tau^{5/2})}{1+\frac{2\alpha}{5}\tau^{3/2}
+\frac{8\pi \alpha}{3\eta_B}\tau^{5/2}},
\end{eqnarray}  
reduces to Eqs. (\ref{LHYBE}) and (\ref{LHYBE2}) in weak-coupling and unitarity limits, respectively.

In the weak-coupling limit, the energy density of a homogeneous superfluid of 
spin-1/2 Fermi gas with an equal population of spin-up and -down fermions is given by \cite{lagfermion,LHYf}
\begin{eqnarray}\label{LHYFE}
{\cal E}_F(n_F,a_F)= \frac{3}{5}E_F\left[ 1+ c_1 \kappa+c_2 \kappa^2+ ...\right],\\
c_1= \frac{10}{9\pi}, \quad c_2 = \frac{4(11-\mbox{ln} 4)}{21 \pi^2},
\quad \kappa=k_Fa_F,
\end{eqnarray}
with Fermi momentum $k_F=(3 \pi^2 n_F)^{1/3}$, Fermi energy 
$E_F = \hbar^2 k_F^2 /2m_F$, and $a_F$ $(<0)$ the $s$-wave scattering length 
corresponding to the scattering between 
spin up-down fermions. In Eq. (\ref{LHYFE}), the first term $3E_F/5$ is the   
density-functional term \cite{BoseFermi_analytical,lagfermion} valid in the BCS 
\cite{bcs} weak-coupling limit {($a_F\to 0^{-}$)} and the following two terms are the 
perturbative LHY corrections \cite{LHYf}  valid for small to medium values of {$|a_F|$}. The unitarity limit of energy density 
is written as \cite{BoseFermi_analytical}
\begin{eqnarray}\label{unitFen}
\lim_{|a_F|\to \infty }{\cal E}_F(n_F,a_F)= 
                        \frac{3}{5}E_F \eta_F.
\end{eqnarray}
The following minimal analytic energy density along the crossover between 
weak-coupling to unitarity regimes \cite{gautam_adhikari}
\begin{eqnarray}\label{enF}
{\cal E}_F(n_F,a_F)= 
\frac{3}{5}E_F\left[1+ \frac{c_1 \kappa +(c_2-2c_1^2)\kappa^2}{1-2c_1 \kappa 
+\frac{(c_2-2c_1^2)\kappa^2}{\eta_F-1}}     
 \right], 
\end{eqnarray}
reduces to Eqs. (\ref{LHYFE}) and (\ref{unitFen}) in the weak-coupling and
unitarity limits, respectively. Again, expression (\ref{enF}) is a Pad\'e-type approximant derived 
from  Eqs. (\ref{LHYFE}) and (\ref{unitFen}).

From Eqs. (\ref{mu}) and (\ref{LHYFE}), the bulk chemical potential of a homogeneous {superfluid} Fermi gas in 
the weak-coupling limit is  
\begin{eqnarray}\label{LHYFMU}
{\cal \mu}_F(n_F,a_F)&\equiv &  \frac{\partial(n_F {\cal E}_F)}{\partial n_F}=
  E_F\left[ 1+d_1  \kappa+d_2 \kappa^2+ ...\right],\\
d_1 &=& \frac{4}{3\pi}, \quad d_2= \frac{4(11-\mbox{ln} 4)}{15 \pi^2},
\end{eqnarray}
and similarly in the unitarity limit from  Eqs. (\ref{mu}) and (\ref{unitFen}), we get for the bulk chemical potential
\begin{eqnarray}\label{unitFmu}
\lim_{|a_F|\to \infty }{\mu}_F(n_F,a_F)=E_F \eta_F.
\end{eqnarray}
The non-perturbative bulk chemical potential along the weak-coupling to unitarity crossover is 
the following Pad\'e-type approximant derived from Eqs. (\ref{LHYFMU}) and (\ref{unitFmu})  \cite{gautam_adhikari}   
\begin{eqnarray}\label{muF}
{\cal \mu}_F(n_F,a_F)  &\equiv&  E_F g(\kappa), \\
g(\kappa)&=&
\left[1+ \frac{d_1 \kappa +(d_2-2d_1^2)\kappa^2}{1-2d_1 \kappa 
+\frac{(d_2-2d_1^2)\kappa^2}{\eta_F-1}}\right].\label{coF}
\end{eqnarray}

For a noninteracting spin-polarized degenerate Fermi gas the energy density is given 
by the so called Fermi pressure term
\begin{eqnarray}\label{dege}
{\cal E}_F(n_F)=& \frac{3}{5}E_F 
\end{eqnarray}
but now with Fermi momentum $k_F=(6 \pi^2 n_F)^{1/3}$ and Fermi energy $E_F = \hbar^2 k_F^2 /2m_F$. 
The difference of a factor of 2 in the Fermi momentum is due to the fact that each energy level can 
now accommodate only one spin-polarized fermion and not two as in the case of the superfluid composed of
an {equal-number} mixture of spin-up and -down fermions. 
In the case of a {\it noninteracting} spin-polarized 
degenerate Fermi gas the bulk chemical potential is 
\begin{eqnarray}\label{degemu}
{\cal \mu}_F(n_F)=E_F  
\end{eqnarray}
and both the energy density and the bulk chemical potential are independent of scattering length. 
Interaction or scattering, {via $s$-partial wave}, between two spin-polarized fermions 
is forbidden by the Pauli exclusion principle.

We consider a Bose/Fermi superfluid in a highly anisotropic 
harmonic trapping potential,  
\begin{equation}
V_i(x,y,z) = \frac{1}{2}{m_i} \omega_i ^2 (\lambda^2 x^2 + \nu^2 y^2 + \gamma^2 z^2)\label{trap_pot},    
\end{equation}
where $\lambda \equiv \omega_x/\omega_i$, $\nu \equiv \omega_y/\omega_i$, and 
$\gamma \equiv\omega_z/\omega_i$ are trap-anisotropy parameters along $x$, $y$ and $z$ 
directions, respectively, and where $\omega_x$, $\omega_y$ and $\omega_z$ are 
angular frequencies of the harmonic traps, respectively, along $x$, $y$ and $z$ axes, 
and $\omega_i$ is a reference frequency that will be defined later. The trap-anisotropy 
parameters $-$ $\lambda, \nu $ and $\gamma$ $-$ are taken to be the  same for bosons 
and fermions, although this restriction can be easily removed.

In the quasi-1D configuration, $\lambda \ll\nu, \gamma$; %the dynamics in these directions can be 
%frozen to be confined 
the 3D wave-function can be approximated as \cite{sala-gll}
\begin{eqnarray} \label{3dwf}
\phi_i^{3D}({\bf r},t) = \phi_{i}^{1D}(x,t)  {\Phi}^{2D}_i(y,z),
\end{eqnarray}
where the dynamics described by the 1D wave function $\phi_{i}^{1D}(x,t) $
is confined along the $x$ direction and 
in the transverse ($y,z$) directions the dynamics is considered frozen in the stationary  harmonic-oscillator 
ground state
\begin{eqnarray}\label{wf2d}
{\Phi^{2D}_i}(y,z) &=& (\pi d_z^2)^{-1/4}   (\pi d_y^2)^{-1/4}e^{-z^2/2 d_z^2}  e^{-y^2/2 d_y^2},  
\end{eqnarray}
  of the 
trapping potential $m_i \omega_i^2( \nu^2 y^2 + \gamma^2 z^2)/2$, where 
$d_z =l_i/\sqrt \gamma$ and  $d_y = l_i/ \sqrt \nu$ with $l_i=\sqrt{\hbar/(m_i\omega_i)}$. 
In this case it is convenient to take {$\omega_i = \omega_x$}.
The bulk chemical potential and bosonic energy density, obtained by conveniently integrating 
out the $y$ and $z$ dependence, will now have the following 1D forms
\cite{sala-gll}
\begin{eqnarray}\label{usp1}
\mu_{i}^{1D}(n_{i}^{1D},a_i)=\int  \mu_i (n_i ,a_i ) |{\Phi}^{2D}_i (y,z)|^2 dy  dz,\\
{\cal E}_{i }^{1D}(n_{i }^{1D},a_i ) = \int  {\cal E}_i (n_i ,a_i ) |{\Phi}^{2D}_i (y,z)|^2 dy  dz,\label{usp2}
\end{eqnarray}
where the 1D density is given by $n_{i}^{1D} = N_i |\phi_{i }^{1D}(x,t)|^2$. 
%with $N_i$ as the respective number of atoms.
 
In the quasi-2D case, $\gamma \gg \lambda,\nu$;  the 3D wave-function can be approximated
as
\begin{eqnarray}\label{wf1d}
\phi_i^{3D} ({\bf r},t) = \phi_{i }^{2D}(x,y,t)  {\Phi}^{1D}_i (z),
\end{eqnarray}
where the dynamics described by the wave function $\phi_{i }^{2D}(x,y,t) $
is confined in the $x$-$y$ plane
and the transverse dynamics along {$z$} direction is considered frozen in the stationary \\ harmonic-oscillator ground state 
\begin{eqnarray}\label{howf1d}
{\Phi}^{1D}_i (z)= (\pi d_z^2)^{-1/4} \exp(-z^2/2 d_z^2), \quad d_z =l_i /\sqrt \gamma,
\end{eqnarray}
  of the trapping potential 
$m_i  \omega_i^2\gamma^2 z^2/2$ with $l_i  =\sqrt{ \hbar/m_i \omega_i}$.
{Here it is convenient to take $\omega_i = \sqrt{\omega_x \omega_y}$.}
The bulk chemical potential and energy density will now have the following 2D forms \cite{sala-gll}
\begin{eqnarray}\label{usp3}
\mu_{i }^{2D}(n_{i }^{2D},a_i )=\int  \mu_i (n_i ,a_i ) |{\Phi}^{1D}_i (z)|^2 dz,\\
{\cal E}_{i }^{2D}(n_{i }^{2D},a_i )=\int  {\cal E}_i (n_i ,a_i ) |{\Phi}^{1D}_i (z)|^2 dz, \label{usp4}
\end{eqnarray}
where the 2D density is now given by $n_{i }^{2D}= N_i |\phi_{i }^{2D}(x,y,t)|^2$.

The dynamical equation  for a  Bose/Fermi superfluid  in the anisotropic trap (\ref{trap_pot}) 
is given by 
\begin{eqnarray}
{\mbox i} \hbar \frac{\partial \phi_{i}^{3D}(x,y,z,t)}{\partial t} &= &
\Big[ -\frac{\hbar^2}{2\beta m_i}\left\{\frac{\partial^2}{\partial x^2}  + \frac{\partial^2}{\partial y^2}  + \frac{\partial^2}{\partial z^2} \right\}
%\nonumber \\
%&  + &
+V_i(x,y,z) + \mu_{i}(n_{i}, a_i)
 \Big]  \phi_{i}^{3D}(x,y,z,t),\nonumber \\
\label{3d}
\end{eqnarray}
 where $\beta =1$ for superfluid bosons and $\beta =4$ for superfluid fermions \cite{BoseFermi_analytical,gautam_adhikari} as well as for degenerate fermions \cite{vonw,tosi}.  Actually, for superfluid fermions the dynamical equation can be written only for a pair of fermions of mass $2m_F$, and the anomalous factor $\beta$ appears after transforming the same  for a single fermion of mass $m_F$.   A similar  factor   has also been  used in the case of degenerate fermions because of phenomenological reasons \cite{vonw,tosi,tosi2,tosi3,tosi4}.

 \section{Quasi-1D and Quasi-2D reduction for bosons}
\label{III}
\subsection{Quasi-1D configuration}
\label{IIIA}
For bosons, in the quasi-1D case, with tight binding in the transverse $y$ and $z$ directions  
($\lambda \ll \nu,\gamma$),
using  Eqs. (\ref{unit}) and (\ref{LHYBE2}) in  Eqs. (\ref{usp1}) and (\ref{usp2}), respectively,
at unitarity, we obtain the following expressions for the bulk chemical potential 
and energy density  
\begin{eqnarray}\label{unit1D}
\lim_{a_B\to \infty} \mu_{B}^{1D}(n_{B}^{1D},a_B)
&=& \frac{\hbar^2}{m_B}\frac{3\eta_B}{5\pi^{2/3}}\delta^{2/3}, \quad \delta = \frac{n_{B}^{1D}}{   d_y d_z},
\\
\lim_{a_B\to \infty} {\cal E}_{B}^{1D}(n_{B}^{1D},a_B) 
&=& \frac{\hbar^2}{m_B}\frac{9\eta_B}{25\pi^{2/3}}  \delta^{2/3}\label{unit1de}.
\end{eqnarray}
Similarly, in the weakly interacting domain ($a_B\to 0$),    Eqs.  (\ref{LHYB}) and (\ref{LHYBE}) 
for the 3D bulk chemical potential and energy density with LHY correction, when inserted into  Eqs. (\ref{usp1}) and (\ref{usp2}), respectively,
yield the following quasi-1D forms for the same
\begin{eqnarray}\label{lhy1D}
{\mu}^{1D}_{B}(n_{B}^{1D},a_B) 
&
=& \frac{2\hbar^2}{m_B}\delta \left[a_B  + \frac{2\alpha}{5\sqrt \pi}
a_B^{5/2}\delta^{1/2}\right],\\
{\cal E}^{1D}_{B}(n_{B}^{1D},a_B) 
&
=& \frac{\hbar^2}{m_B}\delta \left[a_B  + \frac{8\alpha}{25\sqrt \pi}
a_B^{5/2}\delta^{1/2}\right]\label{lhy1de}.
\end{eqnarray}
Equations (\ref{unit1D})-(\ref{unit1de}) and (\ref{lhy1D})-(\ref{lhy1de})
lead to the following Pad\'e-type approximants for the quasi-1D chemical potential and 
energy density along the crossover between weak-coupling and unitarity limits:
%valid from weak coupling to unitarity crossover:
 \begin{eqnarray}
{\mu}^{1D}_{B}(n_{B}^{1D},a_B)  
&=& \frac{{\frac{2\hbar^2}{m_B}}  \delta  \left[ a_B  + \frac{4\alpha}{5\sqrt \pi}
a_B^{5/2}\delta^{1/2} \right] }{1+\frac{2\alpha}{5\sqrt \pi}
a_B^{3/2}\delta^{1/2} +\frac{8\alpha{\pi^{1/6}}}{3\eta_B}a_B^{5/2} \delta^{5/6}}, \label{1DmuB}\\
{\cal E}^{1D}_{B}(n_{B}^{1D},a_B)
&=& \frac{{\frac{\hbar^2}{m_B}}\delta \left[   a_B  + \frac{16\alpha}{25\sqrt \pi}
a_B^{5/2}\delta^{1/2} \right] }{1+\frac{8\alpha}{25\sqrt \pi}
a_B^{3/2}\delta^{1/2} +\frac{16\alpha{\pi^{1/6}}}{9\eta_B}a_B^{5/2} \delta^{5/6}}.
 \label{eq18a}
\end{eqnarray}

If we substitute   the wave function defined by Eqs. (\ref{3dwf})  and (\ref{wf2d}) in the 
dynamical equation (\ref{3d}) and integrate the linear part of this equation 
over the transverse variables $y$ and $z,$ following Ref. \cite{sala-gll},  
and use the reduced
quasi-1D  model bulk chemical potential (\ref{1DmuB}) in the resultant equation, 
we obtain the following
 quasi-1D nonlinear Schr\"odinger equation for relevant dynamics in the  $x$ direction
  for the Bose superfluid  along the crossover, valid between weak coupling 
and unitarity limits 
\begin{eqnarray}
{\mbox i} \hbar \frac{\partial \phi_{B}^{1D}(x,t)}{\partial t} &= &
\Big[ -\frac{\hbar^2}{2m_B}\frac{\partial^2}{\partial x^2}  + 
 \frac{1}{2}m_B \omega_B^2\lambda^2 x^2 
%\nonumber \\
%&  + &
+\mu_{B}^{1D}(n_{B}^{1D}, a_B)
 \Big]  \phi_{B}^{1D}(x,t), 
\label{q1dnlse}
\end{eqnarray}
with the normalization $\int dx | \phi_{B}^{1D}(x,t)|^2 =1$, where ${\mbox i}=\sqrt{-1}$.   
To handle this equation efficiently in a dimensionless form, we express length in units of
$ l_B\equiv \sqrt{\hbar/m_B\omega_B}$, time in units of 
$ m_Bl_B^2/\hbar = \omega_B^{-1}$, density $n_{B}^{1D}\equiv |\phi_B^{1D}|^2$ in units of $l_B^{-1}$, 
and energy in units of $\hbar^2/m_B l_B^2 = \hbar\omega_B$, etc.  Consequently, we obtain the
following dimensionless equation for the relevant dynamics of  the quasi-1D {Bose superfluid} 
along the $x$ direction
 \begin{eqnarray}
{\mbox i} \frac{\partial \tilde{\phi}_{B}^{1D}(\tilde{x},\tilde{t})}{\partial \tilde{t}}  &= &
\Big[ -\frac{1}{2}\frac{\partial^2}{\partial \tilde{x}^2}+ \frac{1}{2} \lambda^2 \tilde{x}^2
%\nonumber \\& +& 
+ \tilde{\mu}_{B}^{1D}(\tilde{n}_{B}^{1D}, \tilde{a}_B) 
\Big]  \tilde{\phi}_{B}^{1D}(\tilde{x},\tilde{t}),
\label{q1dnlse_dimensionless}
\end{eqnarray}
where the quantities with tildes are dimensionless. The transverse dynamics along $y,z$ directions are considered to be frozen in the respective harmonic-oscillator ground state.
Using Eq. {~(\ref{1DmuB})}, we obtain the following  dimensionless 
bulk chemical potential to be used in the reduced quasi-1D  dynamical  equation ~(\ref{q1dnlse_dimensionless}) 
\begin{eqnarray} \label{1db}
\tilde{\mu}_{B}^{1D}(\tilde{n}_{B}^{1D}, \tilde{a}_B) &=& \frac{2 \bar{\delta}\left[\tilde{a}_B  + \frac{4\alpha\sqrt{\bar{\delta}}}{5\sqrt \pi}
\tilde{a}_B^{5/2}\right]}{1+\frac{2\alpha}{5\sqrt \pi}
\tilde{a}_B^{3/2}\bar{\delta}^{1/2} +\frac{8\alpha{\pi^{1/6}}}{3\eta_B}\tilde{a}_B^{5/2} \bar{\delta}^{5/6}},
\end{eqnarray}
where $\bar{\delta} = \sqrt{\nu\gamma}\tilde{n}_{B}^{1D}.$

\subsection{Quasi-2D configuration}

\label{IIIB}

In the quasi-2D case, with tight binding in the transverse $z$ direction ($\lambda, \nu \ll \gamma $), 
using  Eqs. (\ref{unit}) and (\ref{LHYBE2}) for bosons in Eqs.  (\ref{usp3}) and (\ref{usp4}), respectively,
at unitarity, we obtain the following expressions for bulk chemical potential and energy density
 \begin{eqnarray}\label{unit2D}
\lim_{a_B\to \infty} \mu_{B}^{2D}(n_{B}^{2D},a_B)  
&=& \frac{\hbar^2}{m_B}\frac{\sqrt 3 \eta_B}{\sqrt 5\pi^{1/3}}\epsilon^{2/3}, \quad  \epsilon = \frac{n_{B}^{2D}}{d_z},\\
\lim_{a_B\to \infty}{ \cal E}_{B}^{2D}(n_{B}^{2D},a_B)
&=& \frac{\hbar^2}{m_B} \frac{3\sqrt 3 \eta_B}{5\sqrt 5\pi^{1/3}}\epsilon^{2/3}\label{unit2De}. 
\end{eqnarray}
In the weakly interacting domain ($a_B\to 0$),    Eqs.  (\ref{LHYB}) and (\ref{LHYBE}) 
for the 3D bulk chemical potential and energy density with LHY correction, when substituted  into 
Eqs. (\ref{usp3}) and  (\ref{usp4}), respectively, yield the following quasi-2D forms for the same quantities
\begin{eqnarray}\label{lhy2D}
\mu_{B}^{2D}(n_{B}^{2D},a_B) 
&= &\frac{\hbar^2}{m_B}2\sqrt{2\pi} \epsilon \left[  a_B+ \frac{\alpha\sqrt \epsilon a_B^{5/2}}{\sqrt{5\sqrt \pi}}\right],\\
{\cal E}_{B}^{2D}(n_{B}^{2D},a_B) &= &\frac{\hbar^2}{m_B} \sqrt{2\pi} \epsilon \left[  a_B+ \frac{4\alpha\sqrt \epsilon a_B^{5/2}}{5\sqrt{5\sqrt \pi}}\right].\label{lhy2De}
\end{eqnarray}
 Equations (\ref{unit2D}) and (\ref{lhy2D}), and (\ref{unit2De}) and (\ref{lhy2De}), respectively,  
yield the following Pad\'e-type approximants for the quasi-2D chemical potential and energy density
valid along the  weak coupling to unitarity crossover:
 \begin{eqnarray}\label{mu2d}
\mu_{B}^{2D}(n_{B}^{2D},a_B) 
&= &  
\frac{  {\frac{\hbar^2}{m_B}} 2\sqrt{2\pi} \epsilon\left[ a_B+ \frac{2\alpha\sqrt \epsilon a_B^{5/2}}{\sqrt{5\sqrt \pi}} \right] }{1+\frac{\alpha\sqrt \epsilon a_B^{3/2}}{\sqrt{5\sqrt \pi}} + \frac{4 \sqrt {2\pi}\alpha\pi^{1/12}\epsilon^{5/6}a_B^{5/2}}{\sqrt 3\eta_B } },\\
{\cal E}_{B}^{2D}(n_{B}^{2D},a_B)&=&
\frac{{\frac{\hbar^2}{m_B}} \sqrt{2\pi} \epsilon \left[ a_B+ \frac{8\alpha\sqrt \epsilon a_B^{5/2}}{5\sqrt{5\sqrt \pi}} \right] }
{1+\frac{{4}\alpha\sqrt \epsilon a_B^{3/2}}{5\sqrt{5\sqrt \pi}} + \frac{8 \sqrt {2\pi}\alpha\pi^{1/12}\epsilon^{5/6}a_B^{5/2}}{3\sqrt 3\eta_B } }.
\end{eqnarray}

If we substitute   the wave function defined by {Eqs. (\ref{wf1d})  and (\ref{howf1d})} in the 
dynamical equation (\ref{3d}) and integrate the linear part of this equation 
over the transverse {variable $z$}  
and use the reduced
quasi-2D  model bulk chemical potential (\ref{mu2d}) in the resultant equation, 
we obtain the following
 quasi-2D nonlinear Schr\"odinger equation for relevant dynamics in the  $x$-$y$ plane 
  for the Bose superfluid  along the crossover
 \begin{eqnarray}
{\mbox i} \frac{\partial {\phi}_{B}^{2D}( {x},{y},{t})}{\partial {t}} &= &
\Big[ -\frac{1}{2}\frac{\partial^2}{\partial {x}^2}   -\frac{1}{2}\frac{\partial^2}{\partial {y}^2}  + \frac{1}{2}\lambda^2 {x}^2 + 
 \frac{1}{2}\nu^2 {y}^2%\nonumber \\
%&+&
+ {\mu}_{B}^{2D}({n}_{B}^{2D}, {a}_B) 
 \Big] {\phi}_{B}^{2D}({x},{y},{t}),
%\label{q1dnlseb}
\end{eqnarray}
with the normalization $\int d x d y |{\phi}_{B}^{2D}( x, y, t)|^2 =1$.  
Using the dimensionless space and time variables   $\tilde x =x/l_B, \tilde y =y/l_B, \tilde t =t\omega_B$, 
2D density as $\tilde n_B^{2D} = n_B^{2D}l_ B^2$, etc.
the quasi-2D nonlinear Schr\"odinger equation in dimensionless form,  
is then written as 
 \begin{eqnarray}
{\mbox i} \frac{\partial \tilde{\phi}_{B}^{2D}(\tilde{x},\tilde{y},\tilde{t})}{\partial \tilde{t}} &= &
\Big[ -\frac{1}{2}\frac{\partial^2}{\partial \tilde{x}^2}   -\frac{1}{2}\frac{\partial^2}{\partial \tilde{y}^2}  + \frac{1}{2}\lambda^2 \tilde{x}^2 + 
 \frac{1}{2}\nu^2 \tilde{y}^2  %\nonumber \\
%&+&
+ \tilde{\mu}_{B}^{2D}(\tilde{n}_{B}^{2D}, \tilde{a}_B) 
 \Big] \tilde{\phi}_{B}^{2D}(\tilde{x},\tilde{y},\tilde{t}).
\label{q1dnlseb}
\end{eqnarray}
The transverse dynamics along the {integrated $z$} direction is now frozen in the harmonic-oscillator ground state. 
Using  Eq. (\ref{mu2d}),
the dimensionless bulk chemical potential in  Eq. (\ref{q1dnlseb}) along the crossover is given as
\begin{eqnarray}\label{2db}
\tilde{\mu}_{B}^{2D}(\tilde{n}_{B}^{2D},\tilde{a}_B) %&= \frac{ {\sqrt{n_{2D}}}}{ {\sqrt{2 \pi}d_z} }f_{2D}
%(\sigma), \\ &
&=&  
\frac{2\sqrt{2\pi} \bar{\epsilon} \left[  \tilde{a}_B+ \frac{2\alpha\sqrt {\bar{\epsilon}} \tilde{a}_B^{5/2}}{\sqrt{5\sqrt \pi}} \right]   }{1+\frac{\alpha\sqrt {\bar{\epsilon}} \tilde{a}_B^{3/2}}{\sqrt{5\sqrt \pi}} 
+ \frac{4 \sqrt {2\pi}\alpha\pi^{1/12}\bar{\epsilon}^{5/6}\tilde{a}_B^{5/2}}{\sqrt 3\eta_B } },
\end{eqnarray}
where $\bar{\epsilon} = \sqrt{\gamma}\tilde{n}_{B}^{2D}$.

\section{Quasi-1D and Quasi-2D reduction for fermions}
\label{IV}

\subsection{Quasi-1D configuration}
\label{IVA}

Extending the treatment of the previous section to a fully-paired superfluid Fermi gas
at unitarity,  in the quasi-1D case,  Eqs. (\ref{unitFmu}) and (\ref{unitFen}), when substituted into  Eqs. (\ref{usp1})
and (\ref{usp2}), lead to the following expressions for the bulk chemical potential and energy
density of a quasi-1D Fermi superfluid mobile along the $x$ direction with tight binding in transverse 
$y$ and $z$ directions ($ \nu, \gamma \gg \lambda$)
\begin{eqnarray}\label{mufuni}
\lim_{|a_F|\to \infty} \mu_{F}^{1D}(n_{F}^{1D},a_F)%&=&\int dy dz \mu_F(n_F,a_F)|{\Phi}^{2D}_F(y,z)|^2 \nonumber \\ 
&=& \frac{\hbar^2}{m_F}\frac{3\eta_F}{10} \zeta^2, \quad  \zeta   = \frac{(3\pi n_{F}^{1D})^{1/3}}{(d_yd_z)^{1/3}},\\   \label{efuni}
\lim_{|a_F|\to \infty} {\cal E}_{F}^{1D}(n_{F}^{1D},a_F)&= &\frac{\hbar^2}{m_F}
\frac{9\eta_F}{50}\zeta^2, 
\end{eqnarray}
where $\Phi_{F}^{2D}(y,z)$ is now defined by Eq. (\ref{wf2d}) 
with 
$d_z =l_F/\sqrt \gamma = \sqrt{(m_B/m_F)}l_B/\sqrt \gamma, d_y = l_F/ \sqrt \nu = \sqrt{(m_B/m_F)}l_B/ \sqrt \nu$
with $l_F=\sqrt{\hbar/(m_F\omega_F)}$, as we take here $\omega_B=\omega_F$. Equations (\ref{mufuni}) and
(\ref{efuni}) are fermionic counterparts of bosonic equations (\ref{unit1D}) and (\ref{unit1de}).

Similarly, the corresponding equations for the bulk chemical potential and  energy 
density of the paired superfluid fermions in the weak-coupling limit with LHY correction, viz. Eqs. (\ref{LHYFMU}) and (\ref{LHYFE}),
when substituted into Eqs.  (\ref{usp1}) and (\ref{usp2}), lead to the following expressions {for the same quantities in quasi-1D case}
\begin{eqnarray}  \label{muf0}
\mu_{F}^{1D}(n_{F}^{1D}, a_F) &=&   \frac{\hbar^2}{m_F} \frac{3}{10} \zeta^2 \left[1 +   f_1 \bar \zeta 
+ f_2 \bar \zeta^2 +...  \right],\\  \label{ef0}
{\cal E}_{F}^{1D}(n_{F}^{1D}, a_F) &= &  \frac{\hbar^2}{m_F} \frac{9}{50} \zeta^2 \left[1 +  
 g_1 \bar \zeta 
+   g_2 \bar \zeta^2 +...  \right],
\end{eqnarray}
where $f_1=5d_1/6, f_2= 5d_2/7, g_1 =5c_1/6, 
g_2=5c_2/7,$ dimensionless $\bar \zeta = a_F \zeta$.
Equations (\ref{mufuni})-(\ref{efuni}), on the one hand, valid at unitarity, and Eqs. (\ref{muf0})-(\ref{ef0}), on the
other hand, valid in the weak-coupling limit, can be combined together to yield the following Pad\'e-type approximants for the weak coupling to unitarity 
crossover in this case
\begin{eqnarray}\label{xx}
\mu_{F}^{1D}(n_{F}^{1D}, a_F) &=&  \frac{\hbar^2}{m_F} \frac{3}{10} \zeta^2\left[1+ \frac{f_1\bar \zeta +(f_2-2f_1^2)\bar \zeta^2}{1-2f_1\bar \zeta 
+\frac{(f_2-2f_1^2)\bar \zeta^2}{\eta_F-1}}\right], \\ 
{\cal E}_{F}^{1D}(n_{F}^{1D}, a_F) &= &  \frac{\hbar^2}{m_F} \frac{9}{50} \zeta^2\left[1+ \frac{g_1 \bar \zeta +(g_2-2g_1^2)\bar \zeta^2}{1-2g_1\bar \zeta 
+\frac{(g_2-2g_1^2)\bar \zeta^2}{\eta_F-1}}\right].
\end{eqnarray}

For a noninteracting degenerate Fermi gas the quasi-1D chemical potential and energy density  
{obtained by inserting  expressions (\ref{degemu}) and (\ref{dege}) in Eqs. (\ref{usp1}) and (\ref{usp2}), respectively, are }
\begin{eqnarray}\label{df1d}
\mu_{F}^{1D}(n_{F}^{1D}) &=&  \frac{\hbar^2}{m_F} \frac{3}{10} \zeta_1^2, \quad  \zeta_1   = \frac{(6\pi n_{F}^{1D})^{1/3}}{(d_yd_z)^{1/3}} \\
{\cal E}_{F}^{1D}(n_{F}^{1D}) &=&   \frac{\hbar^2}{m_F} \frac{9}{50} \zeta_1^2.
\end{eqnarray}

If we substitute   the wave function defined by Eqs. (\ref{3dwf})  and (\ref{wf2d}) in the 
dynamical equation (\ref{3d}) and integrate the linear part of this equation 
over the transverse variables $y$ and $z,$ following Ref. \cite{sala-gll},  
and use the reduced
quasi-1D  model bulk chemical potential (\ref{xx}) in the resultant equation, 
we obtain the following
 quasi-1D nonlinear Schr\"odinger equation for relevant dynamics in the  $x$ direction
  for the Fermi superfluid  along the crossover, valid between the weak coupling 
and unitarity limits
\begin{eqnarray}
{\mbox i} \hbar \frac{\partial \phi_{F}^{1D}(x,t)}{\partial t} &= &
\Big[ -\frac{\hbar^2}{8m_F}\frac{\partial^2}{\partial x^2}  + 
 \frac{1}{2}m_F \omega_F^2\lambda^2 x^2 
%\nonumber \\
%&  + &
+\mu_{F}^{1D}(n_{F}^{1D}, a_F)
 \Big]  \phi_{F}^{1D}(x,t),
%\label{q1dnlse}
\end{eqnarray}
with the normalization $\int dx | \phi_{F}^{1D}(x,t)|^2 =1$. To write the dimensionless form of this equation we express lengths  in units of $l_B\equiv \sqrt{\hbar/m_B\omega_B}$, 
time in units of $\omega_B^{-1}$, etc. We are using the bosonic length scale for deriving the 
dimensionless equations for fermions for future convenience in writing the same for a 
Bose-Fermi mixture using the same scale.
Then
the dimensionless quasi-1D nonlinear Schr\"odinger equation for a superfluid or a degenerate Fermi gas, for the relevant dynamics along the $x$ direction, using the same
units as used for bosons, is then written as \cite{BoseFermi_analytical,gautam_adhikari,ldim}
\begin{eqnarray}
{\mbox i} \frac{\partial \tilde{\phi}_{F}^{1D}(\tilde{x},\tilde{t})}{\partial \tilde{t}} &=& 
\Big[ -\frac{m_B}{8m_F}\frac{\partial^2}{\partial \tilde{x}^2} + \frac{m_F}{2m_B} \lambda^2 \tilde{x}^2 
%\nonumber \\
%&  + &
+ \tilde{\mu}_{F}^{1D}(\tilde{n}_{F}^{1D}, \tilde{a}_F) 
 \Big]  \tilde{\phi}_{F}^{1D}(\tilde{x},\tilde{t}),
\label{q1dnlsef}
\end{eqnarray}
with the normalization $\int d\tilde{x} | \tilde{\phi}_{F}^{1D}(\tilde{x},\tilde{t})|^2 =1$.
%\sout{In dimensionless form, length is expressed in units of $l_B\equiv \sqrt{\hbar/m_B\omega_B}$, 
%time in units of $\omega_B^{-1}$, etc. We are using the bosonic length scale for deriving the 
%dimensionless equations for fermions for future convenience in writing the same for a 
%Bose-Fermi mixture {\color{red}using the same scale.}} 
The dimensionless crossover formula for the bulk chemical potential {of a superfluid Fermi gas}, 
as obtained from Eq. (\ref{xx}) and  to be used in Eq. (\ref{q1dnlsef}),  is
\begin{eqnarray} \label{1df}
{\tilde \mu}^{1D}_{F}(\tilde n_{F}^{1D}, \tilde a_F) &=&   \frac{3m_B}{10m_F} \widetilde{\zeta}^{2}
\left[1+ \frac{f_1\bar  \zeta +(f_2-2f_1^2)\bar \zeta^2}{1-2f_1\bar  \zeta 
+\frac{(f_2-2f_1^2)\bar \zeta^2}{\eta_F-1}}\right],
\end{eqnarray}
where dimensionless $ \widetilde{\zeta} = \left( 3\pi \tilde n_{F}^{1D}m_F\sqrt{\gamma\nu}/m_B\right)^{1/3}$.
In the case of a spin-polarized degenerate Fermi gas, the dimensionless bulk chemical potential to be used in Eq. (\ref{q1dnlsef}), as obtained from Eq. (\ref{df1d}), can be written as 
\begin{eqnarray} \label{x1}
{\tilde \mu}^{1D}_{F}(\tilde n_{F}^{1D})
= \frac{3m_B^{1/3}}{10m_F^{1/3}}\left( 6\pi \tilde n_{F}^{1D}\sqrt{\gamma\nu}\right)^{2/3}.
\end{eqnarray}

\subsection{Quasi-2D configuration}
\label{IVB}

The  bulk chemical potential and energy density for a quasi-2D Fermi superfluid, confined in the
$x$-$y$ plane with tight binding along the $z$ direction ($\gamma \gg \nu, \lambda$), can be obtained 
using  Eqs. (\ref{usp3}) and (\ref{usp4}). At unitarity, the bulk chemical potential and energy
density, given by Eqs. (\ref{unitFmu}) and  (\ref{unitFen}), respectively, when inserted into
Eqs. (\ref{usp3}) and (\ref{usp4}), lead to the following expressions for these quantities for a quasi-2D Fermi superfluid {mobile along the $x$ and $y$ directions} 
\begin{eqnarray}\label{unesp1}
\lim_{|a_F|\to \infty} \mu_{F}^{2D}(n_{F}^{2D}, a_F)&=& \frac{\hbar^2}{m_F}\frac{\sqrt 3\eta_F}{2\sqrt 5}\chi^2,\\ \label{unesp2}
\lim_{|a_F|\to \infty} {\cal E}_{F}^{2D}(n_{F}^{2D}, a_F)&=& \frac{\hbar^2}{m_F}\frac{3\sqrt 3\eta_F}{10\sqrt 5} \chi^2, 
\end{eqnarray} 
where $\chi =(3 \pi^{3/2}n_{F}^{2D}/d_z)^{1/3}$.
Similarly,  the bulk chemical potential and energy density in weakly interacting domain with LHY correction,
given by   Eqs. (\ref{LHYFMU}) and (\ref{LHYFE}), respectively, when substituted in  Eqs. (\ref{usp3}) and (\ref{usp4}), yield the same quantities for a quasi-2D Fermi superfluid:
\begin{eqnarray}
\mu_{F}^{2D}(n_{F}^{2D}, a_F)
%&=& \frac{\hbar^2}{m_F} \left[\frac{\sqrt 3}{2\sqrt 5}
%{\chi}^2 + \frac{d_1a_F}{2\sqrt 2} {\chi}^3
%+\frac{\sqrt 3 d_2 a_F^2}{2\sqrt 7} {\chi}^{4}\right],\nonumber\\
&=& \frac{\hbar^2}{m_F}\frac{\sqrt 3}{2 \sqrt 5} \chi^2\left[ 1  + h_1\bar  \chi
 + h_ 2\bar  \chi^2  + ...              \right], \label{unesp3} \\
{\cal E}_{F}^{2D}(n_{F}^{2D}, a_F)&=&\frac{3\sqrt 3\hbar^2}{10 \sqrt 5m_F} \chi^2\left[1  +j_1 \bar \chi  +j_ 2\bar  \chi^2  + ...              \right], \label{unesp4}
\end{eqnarray}
where $\bar \chi  =  \chi a_F,  h_1 =\sqrt 5 d_1/\sqrt 6, h_2 =\sqrt 5 d_2/\sqrt 7,\\ 
j_1 =\sqrt 5 c_1/\sqrt 6$, and $j_2 = \sqrt 5 c_2 /\sqrt 7 $.
Equations (\ref{unesp1}) and (\ref{unesp2}), on the one hand, valid at unitarity, and 
 (\ref{unesp3}) and (\ref{unesp4}), on the other hand, valid for weak coupling, can be
combined together to yield the following crossover formulae valid from weak coupling to unitarity 
limit for the quasi-2D bulk chemical potential and energy density 
\begin{eqnarray}
\mu_{F}^{2D}(n_{F}^{2D}, a_F)&=&\frac{\sqrt 3\hbar^2}{2 \sqrt 5m_F} \chi^2\left[1+ \frac{h_1 \bar \chi +(h_2-2h_1^2)\bar \chi^2}{1-2h_1 \bar \chi 
+\frac{(h_2-2h_1^2)\bar \chi^2}{\eta_F-1}}\right], \\
{\cal E}_{F}^{2D}(n_{F}^{2D}, a_F)&=&\frac{3\sqrt 3\hbar^2}{10 \sqrt 5m_F} \chi^2\left[1+ \frac{j_1\bar  \chi +(j_2-2j_1^2)\bar \chi^2}{1-2j_1\bar  \chi 
+\frac{(j_2-2j_1^2)\bar \chi^2}{\eta_F-1}}\right].
\end{eqnarray}

For a noninteracting degenerate Fermi gas, the quasi-2D chemical potential and energy density  
{obtained by inserting  expressions (\ref{degemu}) and (\ref{dege}) into Eqs. (\ref{usp3}) and (\ref{usp4}), respectively,
are} 
\begin{eqnarray}\label{df2d}
\mu_{F}^{2D}(n_{F}^{2D}) &=&  \frac{\hbar^2}{m_F} \frac{\sqrt 3}{2\sqrt 5} \chi_1^2, \quad  \chi_1   = \frac{(6\pi^{3/2} n_{F}^{2D})^{1/3}}{d_z^{1/3}} \\
{\cal E}_{F}^{2D}(n_{F}^{2D}) &=&   \frac{\hbar^2}{m_F} \frac{3\sqrt 3}{10\sqrt 5} \chi_1^2.
\end{eqnarray}

Following the same procedure as above, integrating out the transverse {$z$ variable},
the dimensionless quasi-2D nonlinear Schr\"odinger equation for a superfluid or a degenerate Fermi gas, for the relevant dynamics in  the $x$-$y$  plane, using the same
units as used for bosons, is then written as \cite{BoseFermi_analytical,gautam_adhikari,ldim}
\begin{eqnarray}
{\mbox i}\frac{\partial \tilde \phi_{F}^{2D}(\tilde x,\tilde  y,\tilde  t)}{\partial \tilde  t} &= &
\Big[ -\frac{m_B}{8m_F}\left(\frac{\partial^2}{\partial \tilde  x^2} + \frac{\partial^2}{\partial\tilde  y^2}\right) +\frac{m_F}{2m_B}(\lambda^2\tilde  x^2 + 
 \nu^2\tilde  y^2)
%   \nonumber \\ 
%& +&
+ \tilde \mu_{F}^{2D}(\tilde n_{F}^{2D},\tilde  a_F)
 \Big] \tilde \phi_{F}^{2D}(\tilde x,\tilde y,\tilde t),\nonumber \\
\label{q1dnlsez}
\end{eqnarray}
with the normalization $\int d \tilde x d\tilde y |\tilde \phi_{F}^{2D}(\tilde x,\tilde y,\tilde t)|^2 =1$, here
the dimensionless bulk chemical potential  $ \tilde \mu_{F}^{2D}(\tilde n_{F}^{2D},\tilde  a_F)$ is given by 
\begin{eqnarray}\label{2df}
\tilde \mu_{F}^{2D}(\tilde{n}_{F}^{2D}, \tilde{a}_F)&=&\frac{\sqrt 3m_B}{2 \sqrt 5m_F}\widetilde \chi^2\left[1+ \frac{h_1 \bar \chi +(h_2-2h_1^2)\bar \chi^2}{1-2h_1 \bar \chi 
+\frac{(h_2-2h_1^2)\bar \chi^2}{\eta_F-1}}\right],
\end{eqnarray}
where  dimensionless $\widetilde \chi =(3 \pi^{3/2}{\tilde{n}}_{F}^{2D}\sqrt{\gamma m_F/m_B})^{1/3}$.
For a degenerate Fermi gas Eq. (\ref{q1dnlsez})  remains valid but with the   bulk chemical potential of Eq. (\ref{df2d}) expressed in dimensionless units  
\begin{eqnarray}\label{x2}
\tilde \mu_{F}^{2D}(\tilde{n}_{F}^{2D}) &= & \frac{\sqrt 3\pi m_B^{2/3}}{2 \sqrt 5m_F^{2/3}}(6 {\tilde{n}}_{F}^{2D}\sqrt{\gamma })^{2/3}.
\end{eqnarray}

\section{Model equations for 3D, quasi-1D, and quasi-2D  Bose-Fermi superfluid mixtures}

\label{V}

The dynamical equations for an uncoupled Bose-Fermi superfluid or degenerate mixture (without interspecies interaction) can
readily be written by combining the Bose and Fermi equations of Secs. \ref{III} and \ref{IV}. We will now consider 
a weakly interacting Bose-Fermi mixture. For a Bose-Fermi superfluid mixture, the 3D interspecies energy
density {takes the following form in dimensionless units} 
\begin{eqnarray}
{ \tilde {\cal  E}}_{BF} &=& \frac{2\pi \tilde a_{BF}}{\xi} |\tilde \phi_B^{3D}|^2  |\tilde \phi_F^{3D}|^2,
\end{eqnarray}
where $\xi = m_F/(m_B+m_F)$ and 
$\tilde a_{BF}$ is the Bose-Fermi interspecies scattering length in dimensionless units.
The full 3D model Bose-Fermi equations in dimensionless units are
\begin{eqnarray}\;
{\mbox i}  \frac{\partial\tilde  \phi_B^{3D}({\bf \tilde r},\tilde t)}{\partial \tilde t}&=&
{\Big [}  -\frac{1}{2}\nabla^2 + \frac{\lambda ^2\tilde x^2 + \nu ^2 \tilde y^2 + \gamma^2 \tilde z^2}{2}
+\tilde U_{12} | \tilde \phi_F^{3D}|^2 
\nonumber\\
& +&  \frac{  4\pi \tilde n_B^{2/3}\left(\tau+\alpha \tau^{5/2}\right)}{1+\frac{\alpha}{2}\tau^{3/2}
+\frac{4\pi \alpha}{\eta_B}\tau^{5/2}}
{\Big ]}  \tilde \phi_B^{3D}( {\bf  \tilde r},t),\label{q3db}\\                            
{\mbox i} \frac{\partial \tilde \phi_F^{3D}({\bf r},t)}{\partial  \tilde t}&=&\left [ 
- \frac{m_B}{8m_F} \nabla^2+\frac{m_F}{2m_B} \left( \lambda^2  \tilde x^2+ \nu^2 \tilde y^2 +\gamma^2 \tilde z^2\right) \right.
%\nonumber\\ &+&
+\tilde  U_{21} | \tilde \phi_B^{3D}|^2 
+\frac{m_B(3\pi^2 \tilde n_F)^{2/3}}{2m_F}\nonumber \\
&  \times & \left.\left\{1+ \frac{d_1 \kappa +(d_2-2d_1^2)\kappa^2}{1-2d_1 \kappa 
+\frac{(d_2-2d_1^2)\kappa^2}{\eta_F-1}}\right\} 
\right]  \tilde  \phi_F^{3D}({\bf  \tilde r}, \tilde t),
\label{q3df}
\end{eqnarray}
 with normalization $\int |\tilde\phi_i{^{3D}}({\bf r}, t)|^2d {\bf r} =  1, 
\tilde n_i= N_i |\tilde \phi{_i^{3D}}|^2$,
%= N_i\tilde \rho{_i^{3D}}$,
and where $\tilde U_{12} = {2\pi \tilde a_{BF} }  N_F/{\xi},
\tilde U_{21} = {2\pi \tilde a_{BF} }N_B/{\xi}.$ 

For a degenerate Fermi gas ~(\ref{q3df}) becomes 
\begin{eqnarray}\;                          
{\mbox i} \frac{\partial \tilde \phi_F^{3D}({\bf r},t)}{\partial  \tilde t}&=&\left [ 
- \frac{m_B}{8m_F} \nabla^2+\frac{m_F}{2m_B} \left( \lambda^2  \tilde x^2+ \nu^2 \tilde y^2 +\gamma^2 \tilde z^2\right) \right. \nonumber\\ &+& 
\left. \tilde  U_{21} | \tilde \phi_B^{3D}|^2 
+\frac{m_B(6\pi^2 \tilde n_F)^{2/3}}{2m_F}\right]
 \tilde  \phi_F^{3D}({\bf  \tilde r}, \tilde t).
\label{q3df2}
\end{eqnarray}

For a Bose-Fermi superfluid mixture, quasi-1D and quasi-2D interspecies energy 
densities in the weak-coupling limit in dimensionless units can be written as 
\begin{eqnarray}
{ \tilde {\cal E}}_{BF}^{1D} &=& \frac{2\pi \tilde a_{BF} }{\xi} |\tilde \phi_{B}^{1D}(\tilde x)|^2|
              \tilde \phi_{F}^{1D}(\tilde x)|^2   %\nonumber \\ 
               \int |{\tilde \Phi}^{2D}_B(\tilde y,\tilde z)|^2 |{\tilde \Phi}^{2D}_F(\tilde y,\tilde z)|^2 d\tilde y d\tilde z,\nonumber\\
            &=& 2 \tilde a_{BF} \sqrt{\nu\gamma} |\tilde \phi_{B}^{1D}(\tilde x)|^2|\tilde\phi_{F}^{1D}(\tilde x)|^2,\\
{ \tilde {\cal E}}_{BF}^{2D} &=& \frac{2\pi \tilde a_{BF}}{\xi} |\tilde\phi_{B}^{2D}(\tilde x,\tilde y)|^2|\tilde\phi_{F}^{2D}(\tilde x,\tilde y)|^2 \nonumber
              \int |{\tilde \Phi}^{1D}_B(\tilde z)|^2 |{\tilde \Phi}^{1D}_F(\tilde z)|^2 d\tilde z,\nonumber\\
            &= &\frac{2\sqrt{\pi\gamma} \tilde a_{BF}}{\sqrt{\xi}} |\tilde \phi_{B}^{2D}(\tilde x,\tilde y)|^2|\tilde \phi_{F}^{2D}(\tilde x,\tilde y)|^2,
\end{eqnarray} 
where dimensionless ${\tilde \Phi}^{2D}_{B}(\tilde y,\tilde z), {\tilde \Phi}^{2D}_{F}(\tilde y,\tilde z)$ and ${\tilde \Phi}^{1D}_B(\tilde z), {\tilde \Phi}^{1D}_F(\tilde z)$ are defined as
\begin{eqnarray}
{\tilde \Phi}^{2D}_B(\tilde y,\tilde z)&=& \left(\frac{\gamma\nu }{\pi^2}\right)^{1/4} e^{-\gamma \tilde z^2/2}  e^{-\nu \tilde y^2/2},\\
{\tilde \Phi}^{2D}_F(\tilde y,\tilde z)&= & \left(\frac{\gamma\nu m_F^2}{\pi^2m_B^2}\right)^{1/4}e^{-\gamma m_F \tilde z^2/(2 m_B)}  e^{-\nu m_F \tilde y^2/(2m_B)},\\
{\tilde \Phi}^{1D}_B(\tilde z)&=& \left(\frac{\gamma}{\pi}\right)^{1/4} e^{-\gamma \tilde z^2/2},\\
{\tilde \Phi}^{1D}_F(\tilde z)&=&  \left(\frac{\gamma m_F}{\pi m_B }\right)^{1/4}e^{-\gamma m_F \tilde z^2/(2 m_B)}.
\end{eqnarray} 
Using these interspecies energy densities, the coupled non-linear Schr\"odinger 
equations for the quasi-1D Bose-Fermi superfluid mixture, where the intraspecies 
Bose and Fermi interactions can be varied from weak-coupling to unitarity limits, 
can now be written in dimensionless form as
\begin{eqnarray}\;
{\mbox i}  \frac{\partial \tilde \phi_{B}^{1D}(\tilde x,t)}{\partial t}&=&
{\Big [}  -\frac{1}{2}\frac{\partial^2}{ \partial \tilde x^2}+ \frac{\lambda^2\tilde x^2}{2} +\tilde \mu_{B}^{1D} (\tilde n_{B}^{1D}, \tilde a_B)
 %\nonumber\\ & +&
 +\tilde  U_{12}^{1D} |\tilde \phi_{F}^{1D}|^2 
{\Big ]}  \tilde \phi_{B}^{1D}(\tilde x,t),
\label{q1db}\\
{\mbox i} \frac{\partial \tilde \phi_{F}^{1D}(\tilde x,t)}{\partial t}&=&{\Big [}  
- \frac{m_B \partial^2}{8m_F\partial \tilde x^2}+\frac{m_F}{m_B} \frac{\lambda^2 \tilde x^2}{2} +\tilde  \mu_{F}^{1D} (\tilde n_{F}^{1D}, \tilde a_F) 
 %\nonumber \\  & +&
 +\tilde  U_{21}^{1D} |\tilde \phi_{B}^{1D}|^2 
{\Big ]}  \tilde \phi_{F}^{1D}(\tilde x,t),
\label{q1df}  
\end{eqnarray}
where $\tilde U_{12}^{1D} = 2\tilde a_{BF}N_F \sqrt{\nu\gamma}$ and
$\tilde U_{21}^{1D} = 2\tilde  a_{BF}N_B\sqrt{\nu\gamma}.$
The corresponding equations for quasi-2D system are
\begin{eqnarray}\;
{\mbox i}  \frac{\partial \tilde \phi_{B}^{2D}(\tilde x,\tilde y,t)}{\partial t}&=&
{\Big [}  -\frac{1}{2}\left(\frac{\partial^2}{\partial \tilde x^2}+\frac{\partial^2}{ \partial \tilde y^2}\right) + \frac{\lambda^2\tilde x^2 + \nu^2 \tilde y^2}{2}  \nonumber\\
& +&
\tilde \mu_{B}^{2D} (\tilde n_{B}^{2D},\tilde a_B)
+\tilde  U_{12}^{2D} |\tilde \phi_{F}^{2D}|^2 
{\Big ]}  \tilde \phi_{B}^{2D}(\tilde x,\tilde y,t),\label{q2db}\\
{\mbox i} \frac{\partial \tilde \phi_{F}^{2D}(\tilde x,\tilde y,t)}{\partial t}&=&{\Big [}  
- \frac{m_B}{8m_F} \left(\frac{\partial^2}{\partial \tilde x^2} + \frac{\partial^2}{\partial \tilde y^2}\right)
+\frac{m_F}{2m_B} \left( \lambda^2 \tilde x^2+ \nu^2 \tilde y^2\right)
\nonumber\\ &+&
\tilde  \mu_{F}^{2D} (\tilde n_{F}^{2D}, \tilde a_F) 
+\tilde  U_{21}^{2D} |\tilde \phi_{B}^{2D}|^2 
{\Big ]}  \tilde \phi_{F}^{2D}(\tilde x,\tilde y,t),
\label{q2df}
\end{eqnarray}
where $\tilde U_{12}^{2D} = 2\sqrt{\pi\gamma/\xi}\tilde  a_{BF}N_F$ and $\tilde U_{21}^{2D} = 2\sqrt{\pi\gamma/\xi}\tilde  a_{BF}N_B$. Equations (\ref{q1db})-(\ref{q1df})  and (\ref{q2db})-(\ref{q2df}) with the bulk chemical potentials   $\mu_{i}^{1D} (\tilde n_{i}^{1D}, \tilde a_i)$ and  $\mu_{i}^{2D} (\tilde n_{i}^{2D},\tilde  a_i)$ given by  (\ref{1db}) and 
(\ref{1df}), and  (\ref{2db}) and   (\ref{2df}), respectively, are the principal results of this paper valid for quasi-1D
and quasi-2D Bose-Fermi mixtures. 
For a degenerate Fermi gas, these chemical potentials are given by Eqs.   ({\ref{x1}}) and (\ref{x2}), respectively.
There are no fitting parameters in these equations; the only parameters in these equations are the universal Bertsch parameter \cite{bertsch} at unitarity and the constants of the LHY interaction \cite{LHYbf,LHYb,LHYf}.  Other 
weak coupling to unitarity 
crossover models  for the Bose-Fermi mixture, in addition, have fitting parameters 
\cite{sch-like-eqn,sch-like-eqn1,sch-like-eqn2,sch-like-eqn3,HYD-eqs,HYD-eqs1,BoseFermi_analytical,ldim}, specially for the quasi-1D and quasi-2D configurations.

\section{Numerical results}
\label{VI}

We solve the nonlinear partial differential equations for the Bose-Fermi mixtures of Sec. \ref{V} numerically,
by the split-time-step Crank-Nicolson method employing the imaginary-time propagation \cite{for}, 
using C/Fortran programs \cite{cc,omp}.
We compare the numerical results for reduced 1D densities defined by
\begin{eqnarray}\label{3d1d}
|{\tilde\phi_i}^{3D,\mathrm{R}}(\tilde x)|^2 &\equiv \int d\tilde y d\tilde z |{\tilde\phi_i}^{3D}({\bf \tilde r})|^2,\\
|{\tilde\phi_i}^{2D,\mathrm{R}}(\tilde x)|^2 &\equiv  \int d\tilde y  |{\tilde\phi_i}^{2D}({\tilde x,\tilde y})|^2, \label{2d1d}
\end{eqnarray}
\noindent as appropriate, calculated with quasi-1D, quasi-2D, and full 3D models, 
viz.  (\ref{q1db}-\ref{q1df}),  (\ref{q2db}-\ref{q2df}), and   (\ref{q3db}-{\ref{q3df2}}),
respectively. We will present results for different Bose-Fermi systems of experimental interest,
e.g. $^7$Li-$^6$Li and $^{133}$Cs-$^6$Li.

\begin{figure}[t!]
\begin{center}
\includegraphics[trim = 0mm 0mm 0mm 0mm,clip, width=1\linewidth,clip]{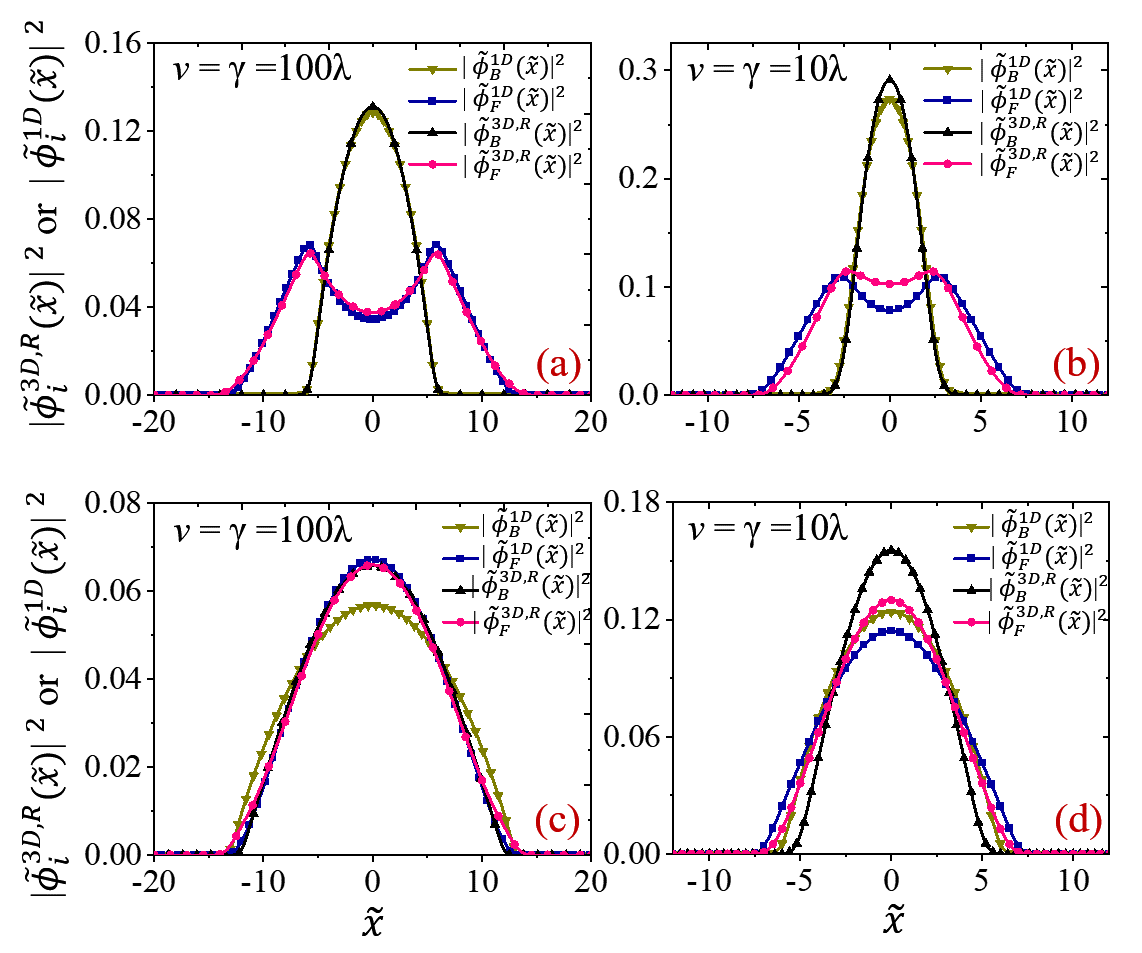}
\caption{(Color online) Quasi-1D  density $|\tilde\phi^{1D}_i(\tilde x)|^2$ of $^{7}$Li-$^{6}$Li
Bose-Fermi superfluid  mixture obtained by solving the quasi-1D  Eqs. (\ref{q1db})-(\ref{q1df}) 
compared with the reduced 1D density $|\tilde\phi_i^{3D,\mathrm{R}}(\tilde x)|^2$ of  Eq.  (\ref{3d1d}) 
obtained by solving the 3D Eqs. 
~(\ref{q3db})-(\ref{q3df}) 
for $N_B = 1000$, $N_F = 100$, {$a_{BF}=100a_0$}, $a_F=-20000a_0$ and with
(a) $\omega_i = \omega_x = 2\pi$ Hz, $\lambda = 1, \nu = \gamma = 100$, $a_{B} = 50 a_0$, 
(b) $\omega_i = \omega_x = 2\pi$ Hz, $\lambda = 1, \nu = \gamma = 10$, $a_{B} = 50 a_0$, 
(c) $\omega_i = \omega_x = 2\pi$ Hz, $\lambda = 1, \nu = \gamma = 100, a_{B} = 500a_0$, and 
(d) $\omega_i = \omega_x = 2\pi$ Hz, $\lambda = 1, \nu = \gamma = 10, a_{B} = 500a_0$. The units of $\tilde {x}$ and densities are $ 38$ $ \mu$m and $ 0.03$ $\mu$m$^{-1}$, respectively.  
}
\label{fig1}
\end{center}
\end{figure}  

\begin{figure}[t!]
\begin{center}
\includegraphics[trim = 0mm 0mm 0mm 0mm,clip, width=1\linewidth,clip]{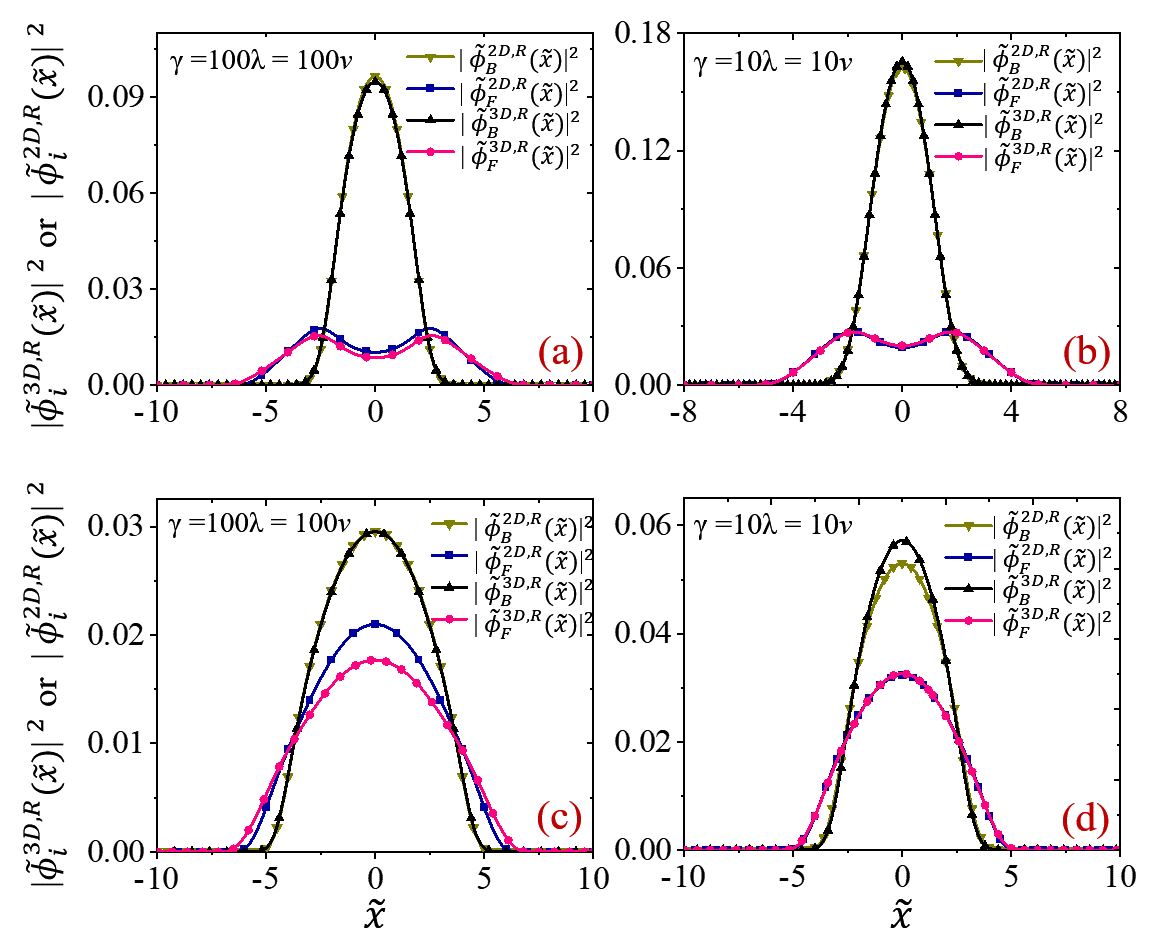}
\caption{ (Color online) Reduced 1D density $|\tilde\phi^{2D,\mathrm{R}}_i(\tilde x)|^2$, viz. Eq.  ~(\ref{2d1d}),
of $^{7}$Li-$^{6}$Li Bose-Fermi superfluid mixture obtained by solving the quasi-2D  Eqs. ~(\ref{q2db})-(\ref{q2df})
and the reduced 1D density $|\tilde\phi_i^{3D,\mathrm{R}}(\tilde x)|^2$ of  Eq. ~(\ref{3d1d}) obtained by
solving 3D Eqs.  (\ref{q3db})-(\ref{q3df}) for $N_B = 1000$, $N_F = 100$, {$a_{BF}=100a_0$}, $a_F=-20000a_0$ and with 
(a) $\omega_i = \omega_x = 2\pi$ Hz, $\lambda =\nu= 1,  \gamma = 100, a_{B} = 50a_0$, 
(b) $\omega_i = \omega_x = 2\pi$ Hz, $\lambda = \nu=1,  \gamma = 10, a_{B} = 50a_0 $,
(c) $\omega_i = \omega_x = 2\pi$ Hz, $\lambda =\nu= 1,  \gamma = 100, a_{B} = 500a_0$, and
(d) $\omega_i = \omega_x = 2\pi$ Hz, $\lambda = \nu=1,  \gamma = 10, a_{B} = 500a_0$.
The units of $\tilde {x}$ and densities are $ 38$ $\mu$m and $ 0.03$ $\mu$m$^{-1}$, respectively.
}
\label{fig2}
\end{center}
\end{figure}  

For a strong Bose-Fermi repulsion, there could be a phase separation between the two components of a
Bose-Fermi mixture \cite{pelster}. A {symmetric} phase separation is characterized by a maximum in the density of one of the components 
coinciding with the local minimum in the density of the other component at the center of the trap. 
Such phase separation can occur in both quasi-1D and quasi-2D configurations, which we elaborate next for a 
superfluid Bose-Fermi mixture. In this paper we use the following values of the Berstch parameters for bosons and fermions, respectively, $\eta_B=4.7, \eta_F=0.415$ \cite{gautam_adhikari}.
The Bose parameter is taken from a fit to a realistic Hartree calculation \cite{dg}, whereas the Fermi parameter is taken from a fit to  realistic Monte-Carlo calculations \cite{MC,gior,MC2} and experiments \cite{expt,expt1,expt2}. 

{\it Trapped quasi-1D Bose-Fermi mixture}: We consider a $^{7}$Li-$^{6}$Li Bose-Fermi 
superfluid binary mixture with $N_B = 1000$, $N_F = 100$, {$a_B=50a_0$}, $a_F=-20000a_0,$  
and {$a_{BF} = 100a_0$}, where $a_0$ is the Bohr radius, in two different trapping potentials: one with 
$\omega_i = 2\pi~{\rm Hz},~\lambda = 1,~\nu = \gamma = 100$ and the other with 
$\omega_i = 2\pi~{\rm Hz},~\lambda = 1,~\nu = \gamma = 10$ in ~(\ref{trap_pot}).
The up-down Fermi-Fermi interaction is taken to be near the unitarity limit.
In this system,  we compare the quasi-1D density $|\phi_{i}^{1D}(\tilde x)|^2$ obtained by solving
the quasi-1D Bose-Fermi equations (\ref{q1db})-(\ref{q1df}) with the reduced 1D density (\ref{3d1d})
 obtained by solving the 3D Bose-Fermi equations (\ref{q3db})-(\ref{q3df}).  
The respective densities are plotted in Fig.  \ref{fig1}  for (a) $\nu=\gamma =100\lambda$,
and (b) $\nu=\gamma =10\lambda.$ Although  there is good agreement between the two densities in both cases, the  agreement between the two 
is better in (a) than in (b), as in the former case a stronger transverse trap creates a more ideal 
quasi-1D confinement than in the latter. In this case, the system shows a phase separation \cite{pelster} between the two components.
For the same set of parameters, if we keep on 
increasing the Bose-Bose repulsion by increasing $a_{B}$,
then the mixture enters into the miscible domain.
 We illustrate this for {$a_{B}=500 a_0$} in two different trapping potentials: 
$\omega_i = 2\pi~{\rm Hz},~\lambda = 1,~\nu = \gamma = 100$ in Fig.  \ref{fig1}(c) and  
$\omega_i = 2\pi~{\rm Hz},~\lambda = 1,~\nu = \gamma = 10$ in Fig. \ref{fig1}(d).
Again, the agreement between the two densities is better in Fig. \ref{fig1}(c) with a stronger transverse trap.
 
\begin{figure}[t!]
\begin{center}
\includegraphics[trim = 0mm 0mm 0mm 0mm,clip, width=\linewidth,clip]{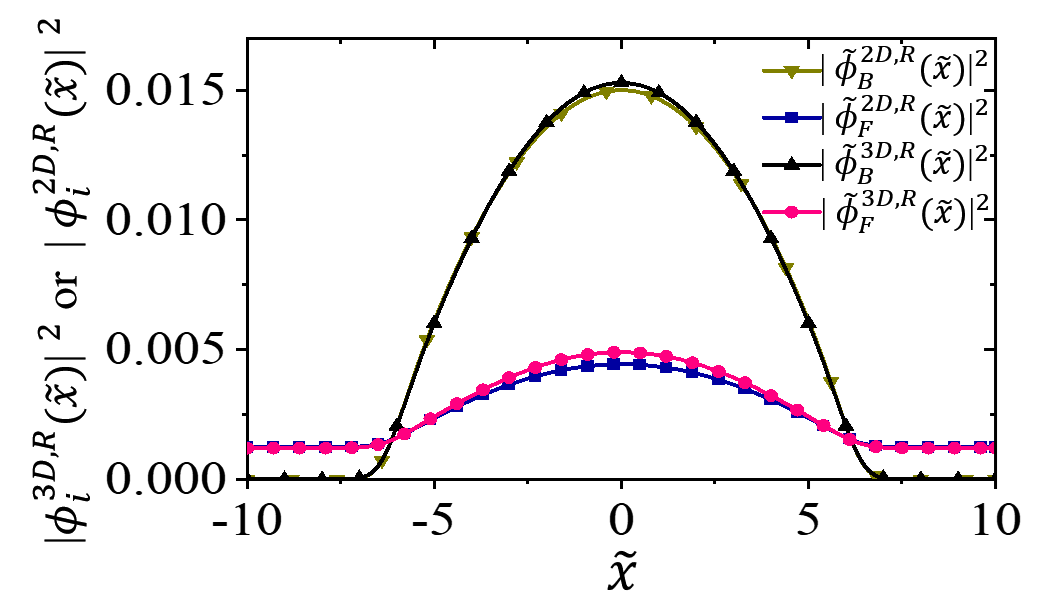}
\caption{(Color online) Reduced 1D density $|\phi^{2D,\mathrm{R}}_i(\tilde x)|^2$, viz. Eq. 
~ (\ref{2d1d}), of bosons and fermions in a quasi-2D Bose-Fermi mixture obtained by 
solving the quasi-2D  equations (\ref{q2db})-(\ref{q2df}) compared with reduced 
1D  density $|\phi^{3D,\mathrm{R}}_i(\tilde x)|^2$, viz. Eq. ~(\ref{3d1d}), obtained by
solving the 3D equations (\ref{q3db})-(\ref{q3df})
for $N_B = 20000$, $N_F = 100$, $a_B=230a_0$, {$a_{BF} = -100a_0$} 
and $a_F = 0^-$. The BEC in the mixture is confined in a harmonic trapping potential 
with $\omega_B = 2\pi$ Hz, $\nu = \lambda = 1, \gamma =100$, whereas Fermi superfluid is confined only along $z$ direction with $\omega_F = 2\pi$ Hz, $\nu = \lambda = 0, 
\gamma =100$. 
The units of $\tilde {x}$ and densities are $ 8.7$ $\mu$m and $ 0.115$ $\mu$m$^{-1}$, 
respectively.
}
\label{fig3}
\end{center}
\end{figure}

{\it Trapped quasi-2D Bose-Fermi mixture}: In this case, we again consider a 
$^{7}$Li-$^{6}$Li Bose-Fermi superfluid mixture with 
$N_B = 1000$,  $N_F = 100$, {$a_B=50a_0$}, $a_F=-20000a_0$, {$a_{BF} = 100a_0$} in two 
different trapping potentials: $\omega_i = 2\pi~{\rm Hz},~\lambda = \nu = 1,~\gamma = 
100$ and $\omega_i = 2\pi~{\rm Hz},~\lambda = \nu = 1,~\gamma = 10$ in Eq.
~(\ref{trap_pot}). In this system, we compare the reduced 1D density $|\phi_{i}^{2D,
{\mathrm{R}}}(\tilde x)|^2$
of Eq.  (\ref{2d1d}) obtained by solving the quasi-2D Bose-Fermi equations (\ref{q2db})-
(\ref{q2df}) with the reduced 1D density $|\phi_{i}^{3D,{\mathrm{R}}}(\tilde x)|^2$
of Eq. (\ref{3d1d}) obtained by solving the 3D Bose-Fermi equations
(\ref{q3db})-(\ref{q3df}). The respective densities are plotted in figure~\ref{fig2}  for 
(a) $\gamma=100\nu =100\lambda$, and (b) $\gamma=10\nu  =10\lambda$.
A minimum of the Fermi density at the center in these cases  signals a phase 
separation between the two components. The agreement between the two densities is good in both cases.

Again for a sufficiently increased Bose-Bose repulsion, the mixture is in the miscible 
domain. We exhibit  the miscible phase for {$a_{B}=500 a_0$} 
in two different trapping potentials:  
$\omega_i = 2\pi~{\rm Hz},~\lambda = \nu = 1,~\gamma = 100$ in Fig. \ref{fig2}(c) and  
$\omega_i = 2\pi~{\rm Hz},~\lambda = \nu = 1,~\gamma = 10$ in Fig. \ref{fig2}(d).
The densities obtained from the quasi-2D equation {are} found to be a better 
approximation to the same densities obtained from the 3D equation, viz. figure
\ref{fig2}, while the densities obtained from the quasi-1D equation may lead to a less 
accurate approximation when compared to the same obtained from the 3D equation, viz. 
figure \ref{fig1}, not only in the present Bose-Fermi 
mixture but also in the case of a single-component BEC \cite{sala-gll}.  
 
\begin{figure}[t]
\begin{center}
\includegraphics[trim = 0mm 0mm 0mm 0mm,clip, width=\linewidth,clip]{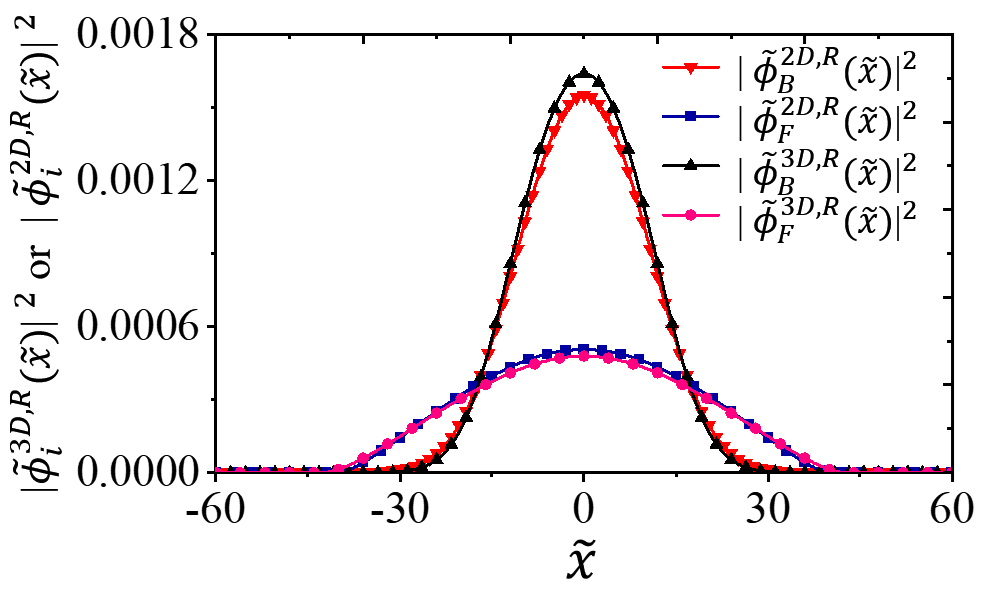}
\caption{ (Color online) Reduced 1D density $|\phi^{2D,\mathrm{R}}_i(\tilde x)|^2$, 
viz. Eq. ~(\ref{2d1d}), of bosons and fermions  in 
a quasi-2D Bose-Fermi mixture obtained by solving the quasi-2D equations ~(\ref{q2db})-
(\ref{q2df}) compared with reduced 1D  density $|\phi^{3D,\mathrm{R}}_i(\tilde x)|^2$, 
viz. Eq. ~(\ref{3d1d}), obtained by solving the 3D equations (\ref{q3db})-(\ref{q3df}) for 
$N_B = 5000$, $N_F = 500$, $a_B = 3a_0, a_{BF} = -100a_0$ 
and $a_F = 0^-$. The Fermi superfluid in the mixture is confined by a harmonic trapping 
potential with $\omega_F = 2\pi$ Hz, $\nu = \lambda = 1, \gamma =100$, whereas the BEC 
is confined only along $z$ direction with $\omega_B = 2\pi$ Hz, $\nu = \lambda = 0, 
\gamma =100$. The units of $\tilde {x}$ and densities are $ 8.7$ $\mu$m and $0.115$ 
$\mu$m$^{-1}$, respectively.
}
\label{fig4}
\end{center}
\end{figure}

In two elegant experiments \cite{Bosonfermi-expt3,Chin_expt_BFsoliton} Desalvo {\em et al.} 
demonstrated that an attractive interspecies interaction in a degenerate Bose-Fermi mixture 
leads to an effective attraction between the bosonic (fermionic) atoms created by the presence 
of the fermionic (bosonic) atoms. This effective attraction may trap bosonic \cite{Chin_expt_BFsoliton} 
(fermionic \cite{Bosonfermi-expt3}) atoms in a trapped Fermi (Bose) {gas} under favorable conditions.
We show that similar trapping is also possible in a superfluid Bose-Fermi mixture for parameters very similar to those used in these experiments and 
we consider this possibility first in the following.

{\it Trapping of a Bose (Fermi) superfluid by a Fermi (Bose) superfluid by fermion- (boson-) mediated interaction}: %Now we discuss the trapping of a BEC (Fermi superfluid) due the effective attraction created by Fermi 
%superfluid (BEC) with an experimentally realistic quasi-2D Bose-Fermi system. 
To study the trapping of free Fermi {superfluid} in a trapped BEC, we consider a $^{133}$Cs-$^6${Li} Bose-Fermi 
superfluid mixture,  where  the spin-up-down fermionic  $^6${Li} atoms are interacting via $s$-wave
scattering length $a_{F}\rightarrow 0^-$. The mixture has an interspecies Feshbach
resonance around 893 Gauss that tunes the interaction between $^6${Li} and $^{133}$Cs \cite{Bosonfermi-expt3}.
Across this resonance, $a_B$ varies from $220a_0$  to  $280a_0$, and we have chosen $a_B = 230 a_0$.
For the Bose gas-mediated trapping of Fermi superfluid, we vary $a_{BF}$ to the requisite values and atoms of
both species are mixed in a certain proportion. In accordance with the experiment by 
Desalvo {\em et al.} \cite{Bosonfermi-expt3}, we consider the harmonically-trapped 
$^{133}$Cs BEC with  angular trap frequencies $\omega_x = \omega_y = 2\pi$ Hz, and 
$\omega_z = 2\pi \times 100$ Hz, whereas the fermionic component is supposed to be
harmonically trapped only along the $z$ direction with angular trap
frequency $\omega_z = 2\pi \times 100$ Hz and free in the $x$-$y$ plane. 
As an example,
we show the trapping of $100$  $^{6}$Li atoms with $a_{F} \rightarrow 0^-$ by a
BEC consisting of $20000$ atoms of $^{133}$Cs  interacting with the Fermi superfluid 
via a Bose-Fermi scattering length {$a_{BF} = -100a_0$}. Due to the Pauli exclusion
principle, only a limited number of fermions can be trapped. In Fig. \ref{fig3} we
plot the reduced 1D densities  $|\phi_{i}^{2D,{\mathrm{R}}}(\tilde x)|^2$ of Eq.  (\ref{2d1d}) of the two components as calculated from a solution 
of the quasi-2D equations   (\ref{q2db}) and (\ref{q2df})   and  compare these with  
the reduced 1D densities $|\phi_{i}^{3D,{\mathrm{R}}}(\tilde x)|^2$ of Eq.  (\ref{3d1d}) 
obtained by solving the 3D Bose-Fermi equations (\ref{q3db})-(\ref{q3df}) in 
Fig. \ref{fig3}.

\begin{figure}[t!]
\begin{center}
\includegraphics[trim = 0mm 0mm 0mm 0mm,clip, width=\linewidth,clip]{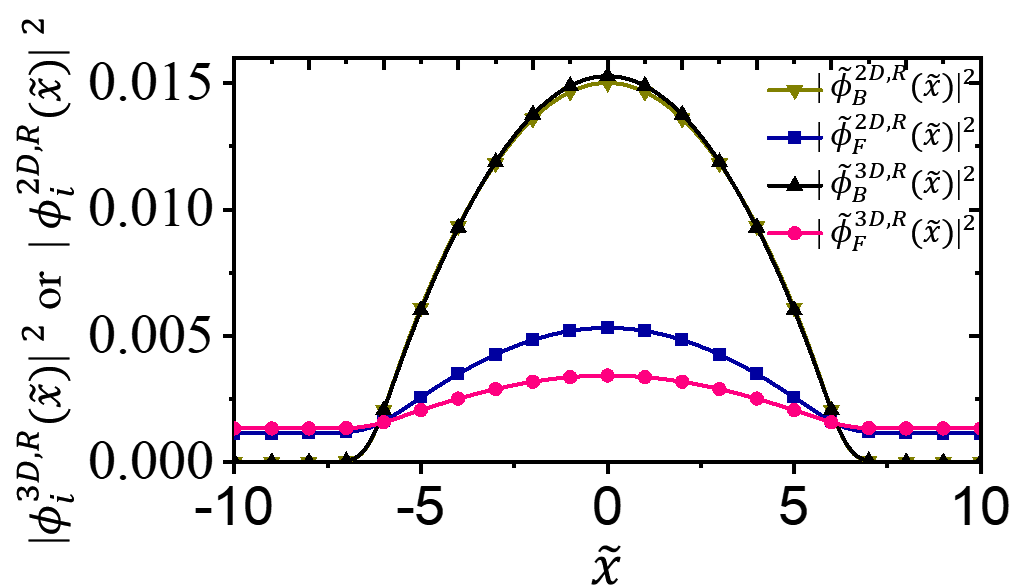}
{
\caption{(Color online) Reduced 1D density $|\phi^{2D,\mathrm{R}}_i(\tilde x)|^2$, viz. ~ (\ref{2d1d}), 
of bosons and fermions in a quasi-2D degenerate Bose-Fermi mixture obtained by solving 
the quasi-2D equations  (\ref{q2db})-(\ref{q2df})
compared with reduced 
1D  density $|\phi^{3D,\mathrm{R}}_i(\tilde x)|^2$, viz. ~(\ref{3d1d}), obtained by 
solving the 3D equations (\ref{q3db}) and (\ref{q3df2})
for  $N_B = 20000$, $N_F = 100$, $a_B=230a_0$, {$a_{BF} = -100a_0$}. The BEC in the mixture is confined by a harmonic trapping potential 
with $\omega_B = 2\pi$ Hz, $\nu = \lambda = 1, \gamma =100$, whereas degenerate Fermi 
gas is confined only along $z$ direction with $\omega_F = 2\pi$ Hz, $\nu = \lambda = 0, 
\gamma =100$. The units of $\tilde {x}$ and densities are $ 8.7$ $\mu$m and $ 0.115$ 
$\mu$m$^{-1}$, respectively.
}
\label{FIGURE3-}}
\end{center}
\end{figure}

\begin{figure}[t]
\begin{center}

\includegraphics[trim = 0mm 0mm 0mm 0mm,clip, width=\linewidth,clip]{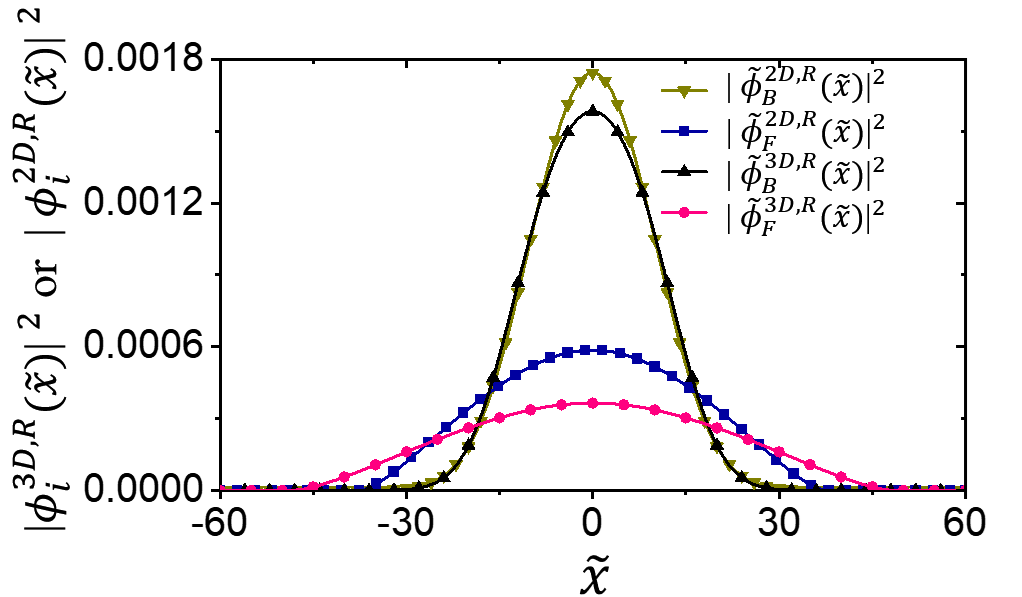}
{
\caption{ (Color online) Reduced 1D density $|\phi^{2D,\mathrm{R}}_i(\tilde x)|^2$, viz. ~(\ref{2d1d}), 
of bosons and fermions in a quasi-2D Bose-Fermi mixture obtained by solving the quasi-2D equations ~(\ref{q2db})-(\ref{q2df})
compared with reduced 1D density $|\phi^{3D,\mathrm{R}}_i(\tilde x)|^2$, viz. ~(\ref{3d1d}), obtained by solving the 3D equations
 (\ref{q3db}) and (\ref{q3df2}) for $N_B = 5000$, $N_F = 500$, $a_B = 3a_0, a_{BF} = -100a_0$. The degenerate Fermi gas in the mixture is confined by a harmonic trapping potential with 
$\omega_F = 2\pi$ Hz, $\nu = \lambda = 1, \gamma =100$, whereas the BEC is confined 
only along $z$ direction with $\omega_B = 2\pi$ Hz, $\nu = \lambda = 0, \gamma =100$. 
The units of $\tilde {x}$ and densities are $ 8.7$ $\mu$m and $0.115$ $\mu$m$^{-1}$, 
respectively.
}
%}
\label{figure4or6}}
\end{center}
\end{figure}

Similarly, to study the trapping of  bosonic atoms in a trapped Fermi superfluid,
we again consider a $^{133}$Cs-$^6${Li} Bose-Fermi superfluid
mixture, where  the spin-up-down fermionic  $^6${Li} atoms are interacting via $s$-wave 
scattering length $a_{F}\rightarrow 0^-$. The fermionic $^6${Li} atoms are harmonically 
trapped with angular frequencies $\omega_x = \omega_y = 2\pi$ Hz, and 
$\omega_z = 2\pi \times 100$ Hz, whereas the bosonic
$^{133}$Cs atoms are harmonically-confined along  the $z$ direction
with angular frequency  $\omega_z = 2\pi \times 100$ Hz and free in the $x$-$y$ plane. 
We show that with attractive interspecies interaction characterized by $a_{BF} = -100a_0$ 
the BEC turns out to be self-trapped in $x$-$y$ plane too as is illustrated in figure \ref{fig4}, where 
we plot the reduced densities (\ref{2d1d}) of the two components as calculated from a solution 
of the quasi-2D equations (\ref{q2db}) and (\ref{q2df}) and compared these with  
the reduced densities $|\phi_{i}^{3D,{\mathrm{R}}}(\tilde x)|^2$ in Eq. ~(\ref{3d1d}) 
obtained by solving the 3D Bose-Fermi equations (\ref{q3db}-\ref{q3df}).
In both Figs. \ref{fig3} and \ref{fig4},
the densities obtained from a solution of quasi-2D model are in {reasonable} agreement with those obtained from the 3D model
for Bose-Fermi superfluid mixtures.

Finally, we consider the trapping of the Bose (Fermi) component by the interaction mediated 
by the Fermi (Bose) component in a $^{133}$Cs-$^6$Li
Bose-Fermi degenerate mixture exactly as in the experiments of 
Desalvo {\em et al.} \cite{Bosonfermi-expt3}. The bosons are in a superfluid state whereas the fermions in a degenerate spin-polarized state. 
We present results for reduced 1D densities of a degenerate Bose-Fermi mixture
in Figs. \ref{FIGURE3-}  and \ref{figure4or6} obtained from a solution  the quasi-2D Bose-Fermi equations (\ref{q2db})-(\ref{q2df})
and compare these with a solution of the 3D Bose-Fermi equations (\ref{q3db}) and (\ref{q3df}) 
using exactly the same parameters as in Figs. \ref{fig3} and \ref{fig4}  for the $^{133}$Cs-$^6$Li
 Bose-Fermi superfluid mixture. For the degenerate mixture, the bulk chemical potentials for fermions  are given by Eq.~(\ref{x2})  
for the quasi-2D case and by {Eq.~(\ref{degemu})} for the 3D case.
If we compare Figs.  \ref{fig3} and \ref{FIGURE3-}, both with $N_B=20000$ and  $N_F=100$,
we note that these systems are essentially bosonic with a small fermionic part. Consequently, the bosonic density in both cases are practically the same with a small change in  the fermionic density.
In Figs.  \ref{fig4} and \ref{figure4or6}, both with $N_B=5000$ and  $N_F=500$, although the number of bosons 
is larger than the number of fermions, the latter is not negligible. Consequently, both densities  have changed in the degenerate Fermi case. The reduced 1D densities obtained from  the quasi-2D model 
in the superfluid Fermi case is a better approximation to the reduced 1D densities obtained from  the 3D 
model, viz.  Fig.  \ref{fig4} 
when compared to the same in the degenerate Fermi gas case,  viz. Fig. \ref{figure4or6}.

\section{SUMMARY}
\label{VII}

We have formulated dynamical equations for quasi-1D and quasi-2D 
superfluid and also degenerate
Bose-Fermi mixtures for the weak coupling to unitarity crossover without any fitting parameters; the only parameters of the 
model are the constants of the beyond-mean-field LHY interactions for bosons and fermions \cite{LHYbf,LHYb,LHYf} 
and the universal Bertsch parameter \cite{bertsch} at unitarity.  We tested these numerically in a few problems 
of superfluid Bose-Fermi mixture confined by quasi-1D and quasi-2D traps. First we considered the problem of 
phase-separated  { and miscible} Bose-Fermi mixture where both components are subject either to 
a quasi-1D trap or a quasi-2D trap. We also considered the problem of self-trapping of one of the components 
in the other trapped component in a quasi-2D Bose-Fermi superfluid as well as in a degenerate Bose-Fermi mixture. 
{Here} the first component is subjected to a quasi-2D trap, whereas the self-trapped second component 
is subjected to a trap only in the $z$ direction. In both cases $-$ the problem of phase separation and of self-trapping 
of one of the components by the other  $-$  we found that the densities obtained from the quasi-1D and quasi-2D model 
equations are in good agreement with the densities obtained from a solution of the full 3D model.

\section*{Acknowledgments}  
 S.G. acknowledges the support of the Science and Engineering Research Board (SERB), Department of Science and 
 Technology, Government of India under the project \\ CRG/2021/002597. 
 S.K.A. acknowledges support by Conselho Nacional de Desenvolvimento Científico e Tecnológico (CNPq), Brazil under grant 301324/2019-0.
 
 %\end{acknowledgments}

\section*{Data Availability}
 
 Data Availability Statement: No Data associated in the ma\-nuscript.


\begin{thebibliography}{99}
\bibitem{Binary-expt1}
G. Modugno, M. Modugno, F. Riboli, G. Roati,  M.
Inguscio,  Phys. Rev. Lett.
{\bf 89}, 190404 (2002).
\bibitem{Binary-expt2}
S. B. Papp, J. M. Pino,  C. E. Wieman, 
Phys. Rev. Lett. {\bf 101}, 040402 (2008).
\bibitem{Binary-expt3}
K. E. Wilson, A. Guttridge, J. Segal,  S. L. Cornish,
Phys. Rev. A {\bf 103}, 033306 (2021).
\bibitem{Binary-expt4}
X. Cui,  Y. Ma, Phys. Rev. Research {\bf 3}, L012027 (2021).
\bibitem{Binary-theo1}
H. Pu,  N. P. Bigelow, Phys. Rev. Lett. {\bf 80}, 1130 (1998).
\bibitem{Binary-theo2}
E. Timmermans, Phys. Rev. Lett. {\bf 81}, 5718 (1998).
\bibitem{Binary-theo3}
P. Ao,  S. T. Chui, Phys.
Rev. A {\bf 58}, 4836 (1998).
\bibitem{Bosonfermi-expt1}
M. Zaccanti, C. D’Errico, F. Ferlaino, G. Roati, M. Inguscio,  
G. Modugno, Phys. Rev. A {\bf 74}, 041605(R) (2006).
\bibitem{Bosonfermi-expt2}
S. Ospelkaus, C. Ospelkaus, L. Humbert, K. Sengstock,  
K. Bongs, Phys. Rev. Lett. {\bf 97}, 120403 (2006).
\bibitem{bfex1}A. Simoni, F. Ferlaino, G. Roati, G. Modugno,  M. Inguscio,
Phys. Rev. Lett. {\bf 90}, 163202 (2003).
\bibitem{bfex2}S. Inouye, J. Goldwin, M. L. Olsen, C. Ticknor, J. L. Bohn,  D. S. Jin,
Phys. Rev. Lett. {\bf 93}, 183201 (2004)
\bibitem{bfex3}F. Ferlaino, C. D’Errico, G. Roati, M. Zaccanti, M. Inguscio, G. Modugno, 
 A. Simoni,
Phys. Rev. A {\bf 73}, 040702(R) (2006).
\bibitem{bfex4}C. A. Stan, M. W. Zwierlein, C. H. Schunck, S. M. F. Raupach,
 W. Ketterle, Phys. Rev. Lett. {\bf 93}, 143001 (2004).
\bibitem{Bosonfermi-expt7}
I. Ferrier-Barbut, M. Delehaye, S. Laurent, A. T. Grier, M. Pierce, 
B. S. Rem, F. Chevy, C. Salomon, Science {\bf 345}, 1035 (2014).
\bibitem{Bosonfermi-expt4}
V. D. Vaidya, J. Tiamsuphat, S. L. Rolston, J. V. Porto,
Phys. Rev. A {\bf 92}, 043604 (2015).
\bibitem{Bosonfermi-expt5}
X.-C. Yao, H.-Z. Chen,  Y.-P. Wu, X.-P. Liu, X.-Q. Wang, X.
Jiang, Y. Deng, Y.-A. Chen,  J.-W. Pan, Phys. Rev. Lett. {\bf 117}, 145301 (2016).
\bibitem{Bosonfermi-expt6}
R. Roy, A. Green, R. Bowler,   S. Gupta,
Phys. Rev. Lett. {\bf 118}, 055301 (2017).
\bibitem{Bosonfermi-expt3}
B. J. DeSalvo, K. Patel, J. Johansen,   C. Chin,
Phys. Rev. Lett. {\bf 119}, 233401 (2017).
\bibitem{Chin_expt_BFsoliton}
B. J. DeSalvo, K. Patel, G. Cai,   C. Chin,
Nature  {\bf 568}, 61 (2019).


\bibitem{pro-cs-li}
S.-K. Tung, C. Parker, J. Johansen, C. Chin, Y. Wang,  P. S. Julienne,
Phys. Rev. A {\bf 87}, 010702(R) (2013).
\bibitem{pro-cs-li-2}
S. K. Tung, K. Jimenez-Garcia, J. Johansen, C. V. Parker,   
C. Chin, Phys. Rev. Lett. {\bf 113}, 240402 (2014).
\bibitem{pro-cs-li-3}
J. Johansen, B. J. DeSalvo, K. Patel,  C. Chin, Nature Phys.
{\bf 13}, 731 (2017).
\bibitem{pro-cs-li-4}
J. Ulmanis, S. Häfner, R. Pires, E. D. Kuhnle, M. Weidem\"uller,   E. Tiemann, New J. Phys. {\bf 17}, 055009 (2015).
\bibitem{pro-cs-li-5}
M. Repp, R. Pires, J. Ulmanis, R. Heck, E. D. Kuhnle,  M.
Weidem\"uller, E. Tiemann, Phys. Rev. A {\bf 87}, 010701(R)
(2013).


\bibitem{Gajda_PRL_BFsoliton} T. Karpiuk, K. Brewczyk, S. Ospelkaus-Schwarzer, K. Bongs, M. Gajda,
 K. Rzazewski, Phys. Rev. Lett. {\bf 93}, 100401 (2004).


\bibitem{adhisol}S. K. Adhikari, Phys. Rev. A {\bf 72}, 053608 (2005).


\bibitem{lagboson}
 F. Dalfovo, S. Giorgini, L. P. Pitaevskii,  S. Stringari,
Rev. Mod. Phys. {\bf 71}, 463 (1999).



\bibitem{gross} E.P. Gross, Nuovo Cimento {\bf 20}, 454 (1961). 
\bibitem{gross1}
L.P. Pitaevskii, Sov. Phys. JETP. {\bf 13}, 451 (1961).
 



\bibitem{lagfermion}
S. Giorgini, L. P. Pitaevskii,  S. Stringari, Rev. Mod. Phys.
{\bf 80}, 1215 (2008).

\bibitem{HYD-eqs}
T. K. Ghosh,  K. Machida, Phys. Rev. A {\bf 73}, 025601 (2006).
\bibitem{HYD-eqs1}
Y. E. Kim,   A. L. Zubarev, Phys. Rev. A {\bf 72}, 011603(R), (2005).


\bibitem{sch-like-eqn}
N. Manini,   L. Salasnich, Phys. Rev. A, {\bf 71}, 033625 (2005).

\bibitem{sch-like-eqn1}
Y. Chen, K.-Z. Zhang, Y.-L. He, Z.-L. Liu,   L. Zhu, 
Int J Theor Phys {\bf 57} 250 (2018).
\bibitem{sch-like-eqn2}
W. Wen, C. Zhao,  X. Ma, Phys. Rev. A {\bf 88}, 063621 (2013).

\bibitem{sle0}S. Gautam, P. Muruganandam, D. Angom,
Phys. Rev. A {\bf 83}, 023605 (2011).


\bibitem{sch-like-eqn3}
J. Yin, Y.-l. Ma,  G. Huang, Phys. Rev. A {\bf 74}, 013609 (2006).
 
\bibitem{boson}S. K. Adhikari,   L. Salasnich,
Phys. Rev. A {\bf 77}, 033618 (2008).



\bibitem{BoseFermi_analytical}S. K. Adhikari,  L. Salasnich, 
Phys. Rev. A {\bf 78},    043616 (2008).


\bibitem{Fermi_sadhan}
S. K. Adhikari, Phys. Rev. A {\bf 77}, 045602 (2008).


\bibitem{gautam_adhikari}
S. Gautam,  S.K. Adhikari, Phys. Rev. A {\bf 100}, 023626 (2019).



\bibitem{wenwen}J. Li, W. Wen, Y. Wang, X. Ma,  H. Li, 
Phys. Lett. A {\bf 410,} 127543 (2021).
\bibitem{wenwen1}
 W. Wen,   H.-j. Li, New J. 
Phys. {\bf 20}, 083044 (2018).
\bibitem{wenwen2}
P. D\'iaz, D. Laroze, A. \'Avila, B. A. Malomed, Commun. Nonlin. Sci. Num. 
Simul. {\bf 70}, 372 (2019).
\bibitem{wenwen3}
H. Sakaguchi,  B. A. Malomed,
Phys. Rev. Research {\bf 2}, 033188 (2020).
\bibitem{wenwen4}
W. Wen, T. Shui, Y. Shan,  C. Zhu, J. Phys. B 
{\bf 48}, 175301  (2015).


\bibitem{LHYbf}
T. D. Lee,  C.N. Yang, 
Phys. Rev. {\bf 105}, 1119 (1957).

\bibitem{LHYb} T. D. Lee, K. Huang,  C.N. Yang, 
Phys. Rev. {\bf 106}, 1135 (1957).

\bibitem{LHYf} K. Huang,  C. N. Yang, Phys. Rev. {\bf 105}, 767 (1957).



\bibitem{sala-gll} L. Salasnich, A. Parola,  L. Reatto, 
Phys. Rev. A {\bf 72}, 025602 (2005). 


\bibitem{bertsch}A. Bulgac,  G. F. Bertsch, Phys. Rev. Lett.
{\bf 94}, 070401 (2005).


\bibitem{mis-im}F. M. Marchetti, Th. Jolicoeur,  M. M. Parish,
Phys. Rev. Lett. {\bf 103}, 105304 (2009).
\bibitem{mis-im1}
S. K. Adhikari, B. A. Malomed, L. Salasnich,  F. Toigo,
Phys. Rev. A {\bf 81}, 053630 (2010). 
\bibitem{mis-im2}
P. Siegl, S. I. Mistakidis,  P. Schmelcher,
Phys. Rev. A {\bf 97}, 053626 (2018). 
\bibitem{mis-im3}
R. Liao, Phys. Rev. Research {\bf 2}, 043218 (2020).
\bibitem{mis-im4}
J. Nisperuza, J. P. Rubio,  R. Avella, J. Phys. Commun {\bf 6}, 025004 (2022).
\bibitem{mis-im5}
S. K. Adhikari,  L. Salasnich,
Phys. Rev. A {\bf 76}, 023612 (2007).

\bibitem{gior}G. E. Astrakharchik, J. Boronat, J. Casulleras,   S. Giorgini,
Phys. Rev. Lett. {\bf 93}, 200404 (2004). 
\bibitem{gior1}
G. E. Astrakharchik, 
R. Combescot, X. Leyronas,  S. Stringari,
Phys. Rev. Lett. {\bf 95}, 030404 (2005).


\bibitem{univ} U. Eismann, L. Khaykovich, S. Laurent, I. Ferrier-Barbut, 
B. S. Rem, A. T. Grier, M. Delehaye, F. Chevy, C. Salomon, L.-C. Ha, 
 C. Chin,  Phys. Rev. X {\bf 6}, 021025 (2016).
 
 \bibitem{univ1}
W. Li,  T.-L. Ho,  Phys. Rev. Lett. {\bf 108}, 195301 (2012). 
 \bibitem{univ2}
P. Makotyn, C. E. Klauss, D. L. Goldberger, E. A. Cornell, 
D. S. Jin,   Nature Phys. {\bf 10}, 116 (2014).
 \bibitem{univ3}
C. Eigen, J. A. P. Glidden, 
R. Lopes, N. Navon, Z. Hadzibabic,  R. P. Smith,
Phys. Rev. Lett. {\bf 119}, 250404 (2017).


%\bibitem{castin} Y.Castin  F.Werner, Lecture Notes in Physics, 
%{\bf 836}, {\bf 127} (2012).  

\bibitem{bcs}J. Bardeen, L. N.  Cooper,  J. R.  Schrieffer,   
Phys. Rev. {\bf 106}, 162 (1957).


\bibitem{vonw} C. F. von Weizs\"acker, Zeits. f\"ur Physik {\bf 96}, 431 (1935).


\bibitem{tosi}P. Capuzzi, A. Minguzzi, M. P. Tosi,
Phys. Rev. A {\bf 68}, 033605 (2003).


\bibitem{tosi2}M. Centelles, M. Guilleumas, M. Barranco, R. Mayol, M. Pi, Laser Phys. {\bf 16}, 360 (2006).

\bibitem{tosi3}D. M. Jezek, M. Barranco, M. Guilleumas, R. Mayol, M. Pi, Phys. Rev. A {\bf 70,}   043630  (2004).

\bibitem{tosi4}M. Guilleumas, M. Centelles, M. Barranco, R. Mayol, M. Pi, Phys. Rev. A {\bf 72,}   053602  (2005).




%\bibitem{as}S. K. Adhikari,  L. Salanich, New J. Phys. {\bf 11}, 023011 (2009).

\bibitem{ldim}S. K. Adhikari, L. Salasnich, New J. Phys {\bf 11}, 023011 (2009).

\bibitem{for} P.  Muruganandam,   S.  K.  Adhikari,  Comput.  Phys.
 Commun. {\bf 180}, 1888 (2009).

\bibitem{cc} D. Vudragovi\'c, I. Vidanovi\'c,
 A. Bala\v z, P. Muruganandam,  S. K. Adhikari, Comput. 
 Phys. Commun. {\bf 183}, 2021 (2012).

\bibitem{omp} L. E. Young-S., P. Muruganandam, S. K. Adhikari,
V. Loncar, D. Vudragovi\'c,  A. Bala\v z, 
Comput. Phys. Commun. {\bf 220}, 503 (2017).


\bibitem{pelster}S. R\"othel, A. Pelster, Eur. Phys. J. B {\bf 59}, 343 (2007).




\bibitem{dg}Y. Ding and C. H. Greene, Phys. Rev. A {\bf 95}, 053602
(2017).

\bibitem{MC} S. Y. Chang, V. R. Pandharipande, J. Carlson, and K. E.
Schmidt, Phys. Rev. A {\bf 70}, 043602 (2004).
 
\bibitem{MC2}
 S. Pilati and S. Giorgini, Phys. Rev. Lett. {\bf 100}, 030401
(2008).



\bibitem{expt}N. Navon, S. Nascimb\`ene, F. Chevy, and C. Salomon,
Science {\bf 328}, 729 (2010).
 \bibitem{expt1}
M. Horikoshi, M. Koashi, H. Tajima, Y. Ohashi, and M. Kuwata-Gonokami, Phys. Rev. X {\bf 7}, 041004 (2017).
\bibitem{expt2}
M. J. H. Ku, A. T. Sommer, L. W. Cheuk, and M.
W.Zwierlein, Science {\bf 335}, 563 (2012).


\end{thebibliography}
\end{document}